\documentclass[a4paper, 12pt, oneside]{article}
\usepackage{jheppub}

\usepackage{graphicx,wrapfig,float,slashed,subcaption,bbold,bm}
\usepackage{amsmath,amssymb,epsfig,graphicx,xcolor}
\usepackage{epstopdf}
\usepackage{booktabs}
\usepackage{ragged2e}
\usepackage{mciteplus}
\usepackage{mathtools}
\allowdisplaybreaks

\usepackage{tikz} 
\usetikzlibrary{shapes,arrows,positioning,automata,backgrounds,calc,er,patterns}
\usepackage{tikz-feynman}
\tikzfeynmanset{compat=1.1.0}

\usepackage{cancel}

\usepackage{makecell}
\usepackage{multirow}
\usepackage{adjustbox}

\usepackage[normalem]{ulem}

\newcommand{\nc}{\newcommand}
\nc{\non}{\nonumber}
\nc{\hc}{\hbox {H.c.}}
\nc{\noi}{\noindent}
\nc{\barx}{\bar{x}}
\nc{\pbarn}{\;\hbox {pb}}
\nc{\fbarn}{\;\hbox {fb}}

\nc{\hsp}{\hspace{0.5cm}}
\nc{\lsp}{\hspace{1cm}}
\nc{\Lsp}{\hspace{2cm}}
\nc{\LLsp}{\lsp\lsp}
\nc{\lra}{\longrightarrow}
\nc{\p}{\prime}
\nc{\sgn}{\text{sgn}}
\nc{\ph}{\varphi}
\nc{\op}{{\cal O}}

\nc{\beq}{\begin{equation}}  \nc{\eeq}{\end{equation}}
\nc{\bea}{\begin{eqnarray}}  \nc{\eea}{\end{eqnarray}}
\nc{\baa}{\begin{array}}     \nc{\eaa}{\end{array}}
\nc{\bit}{\begin{itemize}}   \nc{\eit}{\end{itemize}}
\nc{\ben}{\begin{enumerate}} \nc{\een}{\end{enumerate}}
\nc{\bce}{\begin{center}}    \nc{\ece}{\end{center}}
\nc{\bpm}{\begin{bmatrix}}   \nc{\epm}{\end{bmatrix}}
\nc{\bvt}{\begin{verbatim}}  \nc{\evt}{\end{verbatim}}

\def\lsim{\mathrel{\raise.3ex\hbox{$<$\kern-.75em\lower1ex\hbox{$\sim$}}}}
\def\gsim{\mathrel{\raise.3ex\hbox{$>$\kern-.75em\lower1ex\hbox{$\sim$}}}}

\def\udots{\mathinner{\mkern1mu\raise1pt\vbox{\kern7pt\hbox{.}}\mkern2mu\raise4pt\hbox{.}\mkern2mu\raise7pt\hbox{.}\mkern1mu}}

\def\gev{\;\hbox{GeV}}

\def\dd{\mathrm d}

\newcommand{\Eq}[1]{Eq.~(\ref{#1})}

\newcommand{\bealg}{\begin{equation}\begin{aligned}} 
\newcommand{\eealg}{\end{aligned}\end{equation}} 

\newcommand{\tr}{\mathrm{tr}}

\allowdisplaybreaks

\newcommand\fverb{\setbox\fverbbox=\hbox\bgroup\verb}
\newcommand\fverbdo{\egroup\medskip\noindent%
			\fbox{\unhbox\fverbbox}\ }
\newcommand\fverbit{\egroup\item[\fbox{\unhbox\fverbbox}]}
\newbox\fverbbox

\global\long\def\b#1{\left(#1\right)}%
\global\long\def\s#1{\left[#1\right]}%
\global\long\def\ML{\mathcal{L}}%
\global\long\def\ha{\frac{1}{2}}

\preprint{\begin{flushright}
UT-WI-02-2026\\
\end{flushright}}

\title{Constraints on Loryons in a Two Higgs Doublet Model} 
\author[a]{Can Kilic,}
\author[a]{Sanjay Mathai,}
\author[b,c]{and Taewook Youn}
\affiliation[a]{Theory Group, Weinberg Institute for Theoretical Physics, University of Texas at Austin, Austin, TX 78712, U.S.A.}
\affiliation[b]{Laboratory for Elementary Particle Physics \\ Cornell University, Ithaca, NY 14853, USA}
\affiliation[c]{School of Physics, Korea Institute for Advanced Study \\  Seoul 02455, Republic of Korea}
\emailAdd{kilic@physics.utexas.edu}
\emailAdd{sanjaym@utexas.edu}
\emailAdd{taewook.youn@cornell.edu}

\abstract{We consider Loryons, particles beyond the Standard Model that receive a significant fraction of their masses from electroweak symmetry breaking, in the context of a two Higgs doublet model. Using scalar Loryons in the $[1,1]$, $[1,3]$ (as well as the equivalent $[3,1]$) and the $[2,2]$ representations of the custodial $SU(2)_L \times SU(2)_R$ global symmetry as benchmarks, we study the constraints on the Loryon parameter space, focusing on unitarity, Higgs decay observables, and the absence of Loryon vacuum expectation values. 
We find that while neutral singlet Loryons remain viable for masses up to 700 GeV, representations containing charged scalars are severely constrained by LHC data, particularly as the fraction of mass generated by symmetry breaking increases.}

\begin{document}

\maketitle
\flushbottom


\section{Introduction}
\label{sec:intro}

The experimental discovery of the Higgs boson marked a pivotal step in particle physics. Since then, all the precision measurements of the Higgs boson have confirmed that the elementary particles in the Standard Model (SM) acquire their masses from the vacuum expectation value (VEV) of a weakly coupled scalar field. While we continue to search for new physics at the TeV scale, it is natural to consider the possibility that particles beyond the Standard Model (BSM) may be coupled to the Higgs field as well, and at least a part of their mass may arise from the same VEV. Particles for which the Higgs VEV is the dominant contribution to their masses, termed "Loryons"~\cite{Banta:2021dek}, exhibit non-decoupling behavior, and the associated parameter space is therefore finite and experimentally accessible.

Importantly, Loryons serve as a paradigmatic example of non-decoupling BSM states whose low-energy description requires a Higgs Effective Field Theory (HEFT)~\cite{Feruglio:1992wf,Bagger:1993zf,Burgess:1999ha,Grinstein:2007iv,Alonso:2012px,Espriu:2013fia,Buchalla:2013rka,Brivio:2013pma,Alonso:2015fsp,Alonso:2016oah,Buchalla:2017jlu,Alonso:2017tdy,deBlas:2018tjm,Falkowski:2019tft,Banta:2022rwg,Kanemura:2022txx,Florentino:2024kkf} description rather than the Standard Model Effective Field Theory (SMEFT)~\cite{Weinberg:1979sa,Buchmuller:1985jz,Leung:1984ni,LHCHiggsCrossSectionWorkingGroup:2016ypw,Brivio:2017vri}. By convention, a Loryon has been defined to have more than half of its mass originating from the Higgs VEV. This being the case, there is no well-defined description of a Loryon in the form of a series expansion around the $SU(2)_L \times U(1)_Y$-symmetric point~\cite{Cohen:2020xca,Alonso:2015fsp,Alonso:2016oah,Falkowski:2019tft}.

In Ref.~\cite{Banta:2021dek}, the authors systematically classified all viable scalar and vector-like fermion representations of Loryons coupling to the SM Higgs field. They derived perturbative unitarity bounds on these representations and assessed the constraints from Higgs coupling measurements, precision electroweak observables, and direct collider searches. They concluded that while the vast majority of fermionic Loryons are now excluded -- barring a few exceptional cases -- scalar Loryons transforming in several representations of the extended $SU(2)_{L}\times SU(2)_{R}$ global symmetry group remain experimentally viable, and constitute promising targets for future searches at the LHC.

It is natural to ask how the parameter space of Loryons is affected if one considers an extended scalar sector for electroweak symmetry breaking.
Arguably the most motivated way to extend the scalar sector is to introduce a second doublet of $SU(2)_L$, commonly referred to as a Two Higgs Doublet Model (2HDM). The spectrum of a 2HDM features five physical states: two CP-even scalars ($h$, $H$), one CP-odd pseudoscalar ($A$), and a charged Higgs ($H^\pm$). In this work, we will only consider 2HDM's in which both Higgs doublets acquire a non-vanishing vacuum expectation value. 2HDM's have a far richer phenomenology compared to a single Higgs doublet -- see Refs.~\cite{Branco:2011iw,Wang:2022yhm} for detailed reviews. The 2HDM parameter space can be described by six physical parameters (in addition to the mass and VEV of the light Higgs boson): the masses of the four additional physical states, the mixing angle $\alpha$ between the CP-even states, and the ratio of the vacuum expectation values of the two doublets, $\tan\beta$. 

It is important to note that a 2HDM, where both doublets acquire vacuum expectation values, itself represents a non-decoupling extension of the scalar sector, as it is not possible to make the additional physical states heavy while keeping the masses of the electroweak gauge bosons constant, and the scalar self-couplings perturbative. Unsurprisingly, the 2HDM parameter space is subject to stringent constraints from theoretical consistency requirements, direct collider searches and flavor observables~\cite{Lee:1977eg,Kanemura:1993hm,Akeroyd:2000wc,Barroso:2013awa,Goodsell:2018tti,Goodsell:2018fex,ATLAS:2020zms,Karan:2023kyj,ATLAS:2024lyh}. Theoretically, the scalar potential must be bounded from below to ensure vacuum stability~\cite{Barroso:2013awa}, while scattering amplitudes among the scalar fields must satisfy perturbative unitarity limits at high energies~\cite{Lee:1977eg,Kanemura:1993hm,Akeroyd:2000wc,He:2001tp,Haber:2010bw,Goodsell:2018tti,Goodsell:2018fex,ParticleDataGroup:2024cfk}. Precision electroweak observables, parameterized by the oblique parameters $S$, $T$, and $U$, severely restrict the mass splittings between the heavy neutral ($H, A$) and charged ($H^\pm$) Higgs bosons~\cite{He:2001tp,Haber:2010bw}; in particular, the $T$ parameter limits the breaking of custodial symmetry, favoring spectra where the charged Higgs is nearly degenerate with one of the heavy neutral scalars.

Experimentally, current LHC data on the 125 GeV Higgs boson strongly favors the
``alignment limit'' ($\cos(\beta-\alpha) \to 0$)~\cite{ATLAS:2024lyh}, where
the light scalar $h$ mimics the SM Higgs. Direct collider searches and flavor
observables impose further bounds that vary by model type. For instance, in
Type-II models, $b \to s\gamma$ measurements require the charged Higgs mass
to be $m_{H^\pm} \gtrsim 570$ GeV~\cite{Misiak:2017bgg}, whereas Type-I
models allow for a lighter $H^\pm$ provided $\tan\beta \gtrsim 2$. Other
channels, such as $H/A \to \tau^+\tau^-$ searches~\cite{ATLAS:2020zms},
strongly constrain high-$\tan\beta$ regions in Type-II and Lepton-Specific
models. Despite these limits, significant parameter space remains open,
particularly in Type-I scenarios with lighter scalar masses.

In this paper, we study the interplay between Loryons and the 2HDM. We parameterize the interactions between the Loryons and the 2HDM scalars, and we consider how constraints on the Loryon parameter space are affected by the extended scalar sector. The 2HDM constraints described above generically favor heavier states, but as already mentioned, the 2HDM represents non-decoupling physics so the masses cannot be raised indefinitely. As a result, we expect the viable Loryon parameter space to shrink and eventually vanish as the 2HDM states become heavy and the scalar self-interactions become stronger. In this paper we will work with a benchmark 2HDM spectrum, to be introduced in Section~\ref{sec:twoHiggs} that represents a compromise between these opposing trends, and gives an idea for the maximal extent of the Loryon parameter space, in a 2HDM and arguably in any extended scalar sector for EWSB.

For Loryons coupled to the SM Higgs only, bounds are primarily determined by the Loryon's mass and the strength of two allowed interaction terms between it and the Higgs allowed by custodial symmetry. In contrast, the 2HDM contains more parameters on which the Loryon phenomenology depends.

This dependency leads to three key phenomenological distinctions. First, the presence of additional heavy scalars opens new scattering channels, modifying the theoretical limits from perturbative unitarity. Second, the extended sector introduces new loop contributions to precision observables -- such as the Higgs diphoton decay rate -- which can interfere with the Loryon contributions, thereby shifting the experimentally allowed windows. Finally, the viable Loryon parameter space is no longer static; it varies dynamically with the 2HDM mixing angles, rotating and reshaping the exclusion bounds compared to the standard model case. Since fermionic Loryons are entirely excluded by current data, we focus on the case of scalar Loryons. We find that certain regions of the scalar Loryon parameter space remain compatible with the 2HDM and may be probed by future experimental searches.

The paper is organized as follows. In Section~\ref{sec:oneHiggs}, we begin with a review of Loryons in the presence of a single Higgs doublet as in the SM. In Section~\ref{sec:twoHiggs}, we consider Loryons coupled to a 2HDM and we investigate the constraints of this setup focusing on a few minimal Loryon representations. We conclude in Section~\ref{sec:con}.

\section{Review: Loryons coupled to a single Higgs doublet}
\label{sec:oneHiggs}

In this section we summarize the phenomenology of Loryons in the presence of a single Higgs doublet~\cite{Banta:2021dek}, focusing in particular on the SM quantum numbers of the Loryons, and their interactions with the Higgs field allowed by the symmetries. We also touch upon how integrating out a Loryon requires the use of the HEFT framework. We describe how the need for perturbative unitarity constraints the mass of the Loryons. We follow this with a discussion of additional constraints, arising from electroweak precision measurements and Standard Model Higgs coupling measurements.

\subsection{Representations}

Given the stringent constraints from electroweak precision measurements, particularly on the $T$ parameter which strictly limits isospin breaking, we concentrate on scenarios where the custodial symmetry group $SU(2)_L\times SU(2)_R$ is a good approximate symmetry of the scalar sector, including the Loryons. We therefore assign Loryons, which we wish to have sizable couplings to the Higgs field(s), into complete representations of this global symmetry, ensuring that their contributions to the $T$ parameter vanish at the one-loop level. 

We denote custodial symmetry representations as $[L,R]_X$, following the conventions in Ref.~\cite{Banta:2021dek}. $X$ is common to all the fields inside the representation, which in general contain degrees of freedom from multiple representations of the weak $SU(2)_L$ group, as well as their complex conjugates, each with their own hypercharge $Y$. $X$ and $Y$ are related to each other via $Y = T_R^3 + X$ for each degree of freedom, where the first term on the right hand side denotes the isospin for that degree of freedom, represented by the entries of the generator $T_R^3$. Thus in the standard basis where $T_R^3$ is diagonal, all degrees of freedom in a column of a $[L,R]_X$ representation, written in matrix form, have the same hypercharge.

To make the global $SU(2)_L \times SU(2)_R$ symmetry manifest, 
one can express the SM Higgs field as a bi-doublet:
\beq
\mathcal H = 
\begin{bmatrix} 
\phi^{0*} & \phi^+ \\ -\phi^- & \phi^0
\end{bmatrix}.\label{eq:Hdefined}
\eeq
In unitary gauge, the VEV for the bi-doublet is then given by
\beq
\langle \mathcal H \rangle = \frac{v}{\sqrt2} \mathbb{1}_{2\times2}.
\label{eq:bidoubletvev}
\eeq
In the $[L,R]_X$ notation, this bi-doublet, consisting of the Higgs doublet in the second column ($Y=+1/2$), and the conjugate field in the first column ($Y=-1/2$), corresponds to the $[2,2]_0$ representation. 

In our study we consider Loryons in the lowest representations of custodial symmetry, and we choose the values of $X$ in each case such that there exist Loryon decay channels that are essentially free of collider constraints (mainly because the lightest Loryon is invisible, any decays of heavier Loryons to lighter ones are prompt and only add a few soft mesons to the event, and there are no fractionally or doubly charged states in the spectrum). Thus as far as collider constraints are concerned, we do not have new information to add to what was presented in Ref.~\cite{Banta:2021dek}. Specifically, we focus on the Loryon representations $[1,1]_0$, $[1,1]_1$, $[3,1]_0$ (and the essentially equivalent case $[1,3]_0$), and finally $[2,2]_0$, which contains both a singlet and a triplet of the electroweak group. While we do not study Loryons in higher representations of custodial symmetry in detail, their phenomenology can be studied with the same tools that we use in this paper. Of course, for more exotic Loryon representations, where there may exist charged/colored states with long lifetimes, or with prompt decays to visible states, a more careful study of constraints from direct collider searches needs to be considered. Since most fermionic Loryon candidates are already excluded in models with a single Higgs doublet, in what follows we focus solely on the scalar Loryons.

\subsection{Lagrangian}
We now turn to writing down the most general Lagrangian for these Loryon representations and studying their decomposition into physical states after electroweak symmetry breaking. We remind the reader that we are requiring custodial $SU(2)$ to be a good symmetry in the Loryon sector. We also assume that there is an approximate $\mathbb{Z}_2$ symmetry under which the Loryons are odd and all other fields are even. This is mainly to ensure that the Loryon potential does not have runaway directions. While we demand that all terms in the Lagrangian with $d\le4$ respect this symmetry, we do allow the symmetry to be broken by higher dimensional terms, whose only effect is to allow the Loryons to decay.

Scalar Loryons $\Phi$ belonging to the representaion $[L,R]_X$ transform as $\Phi \to U_L \Phi U_R^\dagger$, where $U_L$ and $U_R$ correspond to the chosen irreducible representations of $SU(2)_L$ and $SU(2)_R$. The relevant terms in the Lagrangian are then given by
\beq
\mathcal L \supset -\frac{M^2}{2^\rho}\, \tr(\Phi^\dagger \Phi) 
- \frac{A}{2^\rho}\, \tr(\Phi^\dagger \Phi)\, \ha \,\tr(\mathcal H^\dagger \mathcal H) 
- \frac{B}{2^\rho}\,2\,\tr(\Phi^\dagger T_{L}^a \Phi T^{\dot a}_{R})\,2\,\tr(\mathcal H^
\dagger T_{2}^a \mathcal H T_{2}^{\dot a}),
\label{eq:sl0lag}
\eeq
with $a, \dot a = 1,2,3$, and $\rho = 0 (1)$ for a complex (real) representation, and with the Higgs bi-doublet in the $[2,2]_0$ representation as defined in Eq.~\eqref{eq:Hdefined}.
Note that the $B$ term, which involves a sum over the generators of the custodial group, is only defined if neither $L$ nor $R$ are singlet representations - otherwise this term is taken to be zero. The generators are cannonically normalized by
\beq
\tr(T^a_{L} T^b_{L}) = \frac{\mathrm{dim}(L)~C_2(L)}{3} \delta^{ab},
\eeq
and the same for the $R$ representation, where the quadratic Casimir $C_2$ is given by
\beq
C_2(L) = \frac{1}{4} (\mathrm{dim}(L) + 1) (\mathrm{dim}(L) - 1),
\label{eq:C2}
\eeq
and again the same for $R$. It is easy to see based on Eq.~\eqref{eq:bidoubletvev} that the Higgs VEV breaks $SU(2)_{L}\times SU(2)_{R}$ to the diagonal $SU(2)_{V}$ subgroup. The $L \times R$ components of $\Phi$ are then decomposed into a direct sum of $SU(2)_V$ irreps:
\beq
\Phi \to \underset{V \in \mathcal V}{\oplus} \phi_V,
\eeq
where $\phi_V$ is $V$-dimensional with 
\beq
\mathcal V = \bigg\{ L+R-1, L+R-3, \cdots, |L-R|+1 \bigg\}.
\label{eq:vdecom}
\eeq

In unitary gauge, \Eq{eq:sl0lag} becomes
\beq
\mathcal L \supset -\frac{1}{2^\rho} \sum_{V\in\mathcal V}\phi_V^\dagger \left\{ M^2 + \frac12 \lambda_V (v+h)^2 \right\} \phi_V,
\label{eq:sl0lagew}
\eeq
which yields the mass spectrum of the scalar Loryons. Here  $\lambda_V$ is given in terms of the couplings in Eq~\eqref{eq:sl0lag} by
\beq
\lambda_V = A + B[C_2(L)+C_2(R) - C_2(V)].
\eeq

Applying these general results to the representations of interest, we find the following decomposition and mass spectra:
\begin{itemize}
    \item {\bf$[1,1]$ Representation ($L=1, R=1$)}: Note that the $B$-term is absent. The decomposition rule in \Eq{eq:vdecom} yields a single irreducible representation with dimension $V=1$, with the coupling:
    $$\lambda_{1} = A.$$
    \item {\bf$[1,3]$ ($[3,1]$) Representations ($L=1, R=3$ ($L=3, R=1$))}: Once again, the $B$-term is absent. These decompose into a single custodial triplet with $V=3$, with the coupling:
    $$\lambda_{3} = A.$$
    \item {\bf$[2,2]$ Representation ($L=2, R=2$)}: This representation decomposes into a singlet ($V=1$) and a triplet ($V=3$) under $SU(2)_V$. With $C_2(2) = 3/4$, the common term is $C_2(L)+C_2(R) = 3/2$. The couplings for the distinct physical states are: \\
    For the Singlet ($V=1$): Since $C_2(1)=0$,
    $$\lambda_1 = A + B\left[\frac{3}{2} - 0\right] = A + \frac{3}{2}B$$
    For the Triplet ($V=3$): Since $C_2(3)=2$,
    $$\lambda_3 = A + B\left[\frac{3}{2} - 2\right] = A - \frac{1}{2}B$$
    The $A$ parameter contributes to the overall mass scale as in the previous cases, while the $B$ parameter is responsible for generating the mass splitting between Loryon states.
\end{itemize}

\subsection{HEFT Condition}
What makes the Loryon model particularly interesting is the existence of a parameter space where HEFT is the only valid effective field theory description, while SMEFT does not provide a valid expansion. This region can be identified by deriving the leading-order effective Lagrangian using the functional methods developed in Ref.~\cite{Cohen:2020xca}.
For a scalar sector with any custodial irrep scalar Loryon, we have in arbitrary gauge \cite{Banta:2021dek}
\beq
\mathcal L_\mathrm{eff} \supset \frac{1}{2^\rho (4\pi)^2} \sum_{V\in\mathcal V} V \left\{ \frac{m_V^4(H)}{2}\left[\log\frac{\mu^2}{m_V^2(H)} + \frac{3}{2} \right] + \frac{\lambda_V^2}{6m_V^2(H)}\frac{(\partial|H|^2)^2}{2} + \mathcal O(\partial^4) \right\},
\label{eq:efflagscl}
\eeq
where $H$ denotes the SM Higgs doublet and
\beq
m_V(H)^2 = M^2 + \lambda_V |H|^2.
\eeq
The first term in \Eq{eq:efflagscl} is the well known Coleman-Weinberg potential, with the second term being a generalization of the one-loop effective lagrangian to higher order in derivatives as introduced in Ref.~\cite{Cohen:2020xca}. Moreover, we can shift to unitarity gauge by simply replacing $2|H|^2$ with $(v+h)^2$, where $h$ is the physical Higgs boson. Depending on whether one uses SMEFT or HEFT, one would expand Eq.~\eqref{eq:efflagscl} in $H$ or in $h$. A SMEFT description of \Eq{eq:efflagscl} is acquired by expanding $m_V(H)$ about $ H=0$ by treating it as $m_V^2(H) = M^2\left(1 + \frac{\lambda_V|H|^2}{M^2}\right)$. For SMEFT to provide a reliable description of low-energy observables, this expansion must converge when evaluated at the electroweak symmetry-breaking vacuum.
This requirement translates to the condition
\beq
\frac{\lambda_V v^2}{2M^2} < 1
\eeq
for all $V \in \mathcal V$, establishing the necessary matching condition on HEFT
\beq
f_\mathrm{max} \ge \frac12,
\label{eq:fmax}
\eeq
where $f_\mathrm{max} \equiv \max_{V\in\mathcal V} f_V$ with 
\beq
f_V \equiv \frac{\lambda_V v^2 / 2}{M^2 + \lambda_V v^2 / 2}.
\label{eq:fvsc}
\eeq
Therefore, if more than half of a scalar Loryon's mass-squared comes from electroweak symmetry breaking, the HEFT description must be employed, resulting in the effective potential of \Eq{eq:efflagscl}, expanded as a series around $h=0$.

\subsection{Unitarity Bounds}
\label{sec:ub0}
By definition, Loryons derive a significant portion of their mass from the Higgs VEV. As the Loryon’s coupling to the Higgs grows with its mass, the size of this coupling, and in connection the Loryon's mass is constrained by unitarity violation from scattering processes involving the Higgs field. 

Let us consider a matrix of $2\to2$ scattering amplitudes $\left[\mathcal M(i \to f)\right]$, written in a basis where the initial state $i$ and a final state $f$ range over all possible combinations of Loryons and the Higgs boson, evaluated at the same value of $s=(p_1+p_2)^2$ and CM scattering angle $\theta^*$. We define ${\mathcal P}$ as the region of phase space where no elements of $[{\mathcal M}]$ contain diagrams close to poles, adopting the same conventions as in reference~\cite{Goodsell:2018tti}. For instance, ${\mathcal P}$ only includes values of $s$ for which $|1-\frac{s}{m^2}|>0.25$ for any particles with mass $m$ appearing in an $s$-channel diagram. The restriction on $\theta^*$ is more involved, but is essentially aimed at avoiding poles in $t$ or $u$-channel diagrams. Based on partial wave considerations, one can derive constraints on the $J$-th partial wave amplitudes~\cite{Goodsell:2018tti}
\beq
-i ([a_J] - [a_J]^\dagger) \geq [a_J] [a_J]^\dagger, 
\eeq
where $[a_J]$ is a normal matrix obtained from the partial wave decomposition of the matrix $[\mathcal M]$. In particular, the zeroth partial wave is given by
\beq
[a_0] \equiv \frac{1}{32\pi} \sqrt{\frac{4 |\mathbf{p}_i| |\mathbf{p}_f|}{2^{\delta_i + \delta_f}s}} \int^1_{-1} \dd\cos\theta^{*}~[\mathcal M].
\eeq
The factor $\delta_i$/$\delta_f$ is 1 when the initial/final particles are identical, and 0 otherwise. 

The unitarity bounds are strongest near kinematic threshold where propagator effects are enhanced - in particular due to $t$ and $u$-channel propagators going on-shell in the forward limit. That being the case, we focus our attention on $[a_0]$, since higher partial waves are velocity suppressed near threshold. For $a^k_0$ being the eigenvalues of the matrix $[a_0]$, the unitarity of the S-matrix then imposes the bound
\beq
\max_{s \in {\mathcal P},k}|\Re(a^k_0)| \le \frac{1}{2} .
\label{eq:ub}
\eeq
where $\Re$ denotes the real part. 

Note that these constraints do not point to a definitive breakdown of unitarity. Nevertheless, they serve as a valuable guide for identifying the range in the Loryon parameter space in which perturbation theory remains applicable. Beyond this unitarity-constrained range, Loryon couplings become large and bound states may form, which fall outside the scope of the theoretical framework discussed here.

\begin{figure}[h!]
    \centering
    \includegraphics[width=0.7\linewidth]{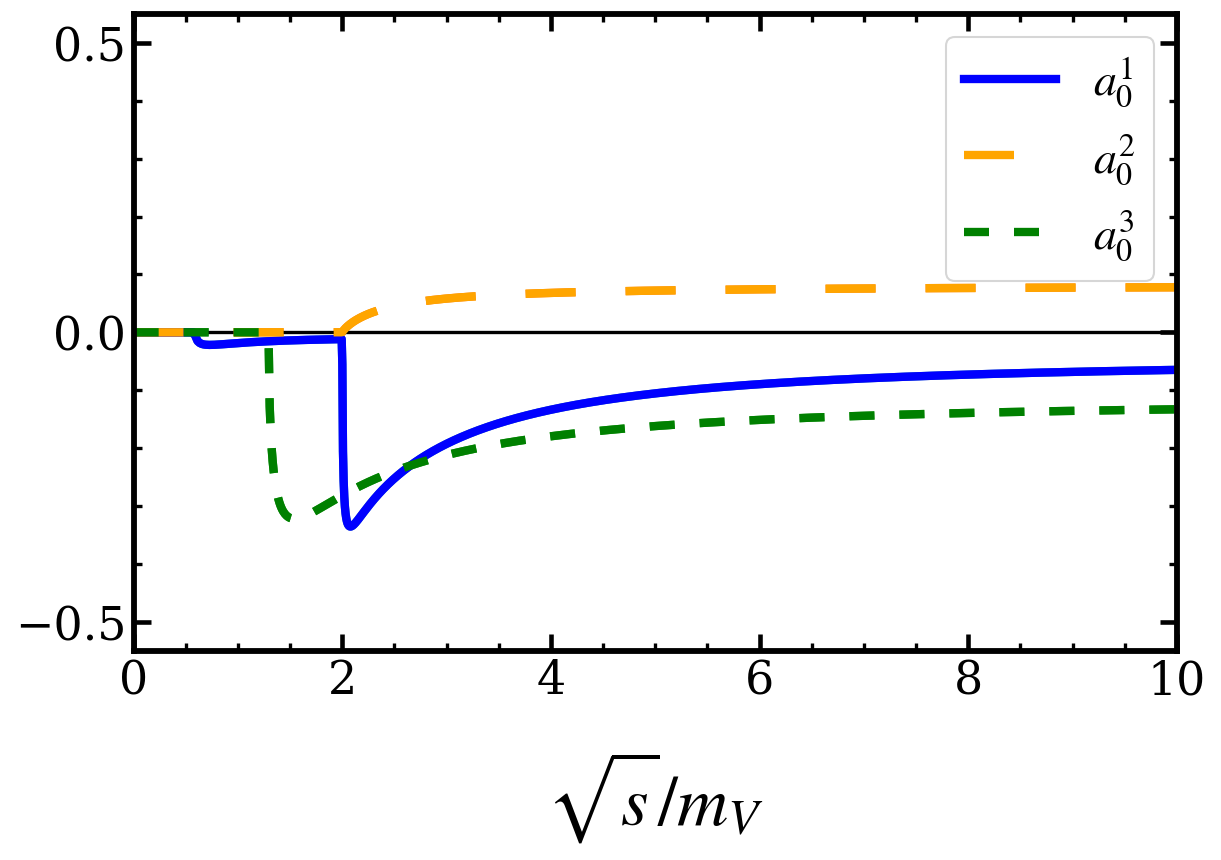}
    \caption{The values of the three eigenvalues of the $[a_0]$ matrix as a function of to the center-of-frame energy $\sqrt s$, shown for $f_\phi = 1$ and Loryon mass $m_V = 425~\text{GeV}$.}
    \label{fig:enter-label}
\end{figure}

As the simplest example of unitarity constraints, we consider the interactions given in \Eq{eq:sl0lag} (or (\ref{eq:sl0lagew}) after electroweak symmetry breaking), involving a neutral singlet Loryon field $\phi$ and the physical Higgs boson $h$. The corresponding Lagrangian reads
\begin{equation}
    \ML \supset -\ha f_\phi \frac{2m_\phi^2}{v} h \, \phi^2 -\frac{1}{4} f_\phi \frac{2m_\phi^2}{v^2} h^2 \, \phi^2 - \frac{1}{24}\lambda_4 \phi^4 -  \frac{1}{6}\frac{3m_h^2}{v} h^3 - \frac{1}{24} \frac{3m_h^2}{v^2}h^4 \, 
    \label{eq:lhphi}
\end{equation} 
where $f_\phi$ denotes the fraction of the singlet Loryon mass arising from the Higgs as defined in Eq.~\eqref{eq:fvsc}. We consider the tree-level $2 \to 2$ scattering processes derived from this Lagrangian, with initial and final states of $\phi\phi$, $hh$ and $\phi h$. 
$[{\mathcal M}]$ and therefore also $[a_0]$ can in general be put into block-diagonal form, as the $\mathbb{Z}_2$ symmetry of the Loryons, as well as $CP$ impose superselection rules, setting the scattering amplitude for many of the initial/final state combinations to zero. For this particular case, $[a_0]$ for $\phi$–$h$ scattering can be written in the form:
\begin{equation}
\begin{bmatrix} \scalebox{2}{$a_0$} \end{bmatrix} \qquad \scalebox{2}{$=$} \qquad 
\begin{bmatrix}
    \begin{tikzpicture}[baseline=(a.base)]
        \begin{feynman}[small]
            \vertex [blob] (a) {};
            \vertex [above left=of a] (b) {$\phi$};
            \vertex [below left=of a] (c) {$\phi$};
            \vertex [above right=of a] (d) {$\phi$};
            \vertex [below right=of a] (e) {$\phi$};
            \diagram* { (b) -- (a) -- (c), (e) -- (a) -- (d) };
        \end{feynman}
    \end{tikzpicture}
    & 
    \begin{tikzpicture}[baseline=(a.base)]
        \begin{feynman}[small]
            \vertex [blob] (a) {};
            \vertex [above left=of a] (b) {$\phi$};
            \vertex [below left=of a] (c) {$\phi$};
            \vertex [above right=of a] (d) {$h$};
            \vertex [below right=of a] (e) {$h$};
            \diagram* { (b) -- (a) -- (c), (e) -- (a) -- (d) };
        \end{feynman}
    \end{tikzpicture}
    &
    \scalebox{2}{$0$}
    \\ 
    \begin{tikzpicture}[baseline=(a.base)]
        \begin{feynman}[small]
            \vertex [blob] (a) {};
            \vertex [above left=of a] (b) {$h$};
            \vertex [below left=of a] (c) {$h$};
            \vertex [above right=of a] (d) {$\phi$};
            \vertex [below right=of a] (e) {$\phi$};
            \diagram* { (b) -- (a) -- (c), (e) -- (a) -- (d) };
        \end{feynman}
    \end{tikzpicture}
    & 
    \begin{tikzpicture}[baseline=(a.base)]
        \begin{feynman}[small]
            \vertex [blob] (a) {};
            \vertex [above left=of a] (b) {$h$};
            \vertex [below left=of a] (c) {$h$};
            \vertex [above right=of a] (d) {$h$};
            \vertex [below right=of a] (e) {$h$};
            \diagram* { (b) -- (a) -- (c), (e) -- (a) -- (d) };
        \end{feynman}
    \end{tikzpicture}
    & 
    \scalebox{2}{$0$}
    \\
    \scalebox{2}{$0$} & \scalebox{2}{$0$} &
    \begin{tikzpicture}[baseline=(a.base)]
        \begin{feynman}[small]
            \vertex [blob] (a) {};
            \vertex [above left=of a] (b) {$h$};
            \vertex [below left=of a] (c) {$\phi$};
            \vertex [above right=of a] (d) {$h$};
            \vertex [below right=of a] (e) {$\phi$};
            \diagram* { (b) -- (a) -- (c), (e) -- (a) -- (d) };
        \end{feynman}
    \end{tikzpicture}
    \\
\end{bmatrix}.
\label{eq:a0m1}
\end{equation}
The eigenvalues of $[a_0]$ are plotted in Figure~\ref{fig:enter-label} as a function of the center-of-mass energy $\sqrt{s}$, with $f_\phi = 1$ and the Loryon mass is taken as $m_V =425~\text{GeV}$.  The $J=0$ partial wave is maximized at values of $\sqrt{s}$ near threshold, and not in the high energy limit as shown in Ref.~\cite{Goodsell:2018tti}. Near threshold, the largest contribution comes from the $t$-channel exchange in the process $\phi\phi\rightarrow\phi\phi$, thus this is the source of the unitarity constraint.

While the quartic Loryon self-couplings are also bounded by unitarity, they are not as strongly constrained as the Loryon-Higgs couplings. Consequently, their values do not play a role in our analysis.

\subsection{Electroweak Precision Measurements}
\label{ssec:ewpm}
Electroweak precision measurements can significantly constrain the Loryons, which carry SM electroweak quantum numbers. As the primary interactions of the new particles with the SM occur through the Higgs and gauge bosons, the oblique parameter framework \cite{Lynn:1985fg,Peskin:1991sw} can be used to study the most stringent constraints.
There are seven extended electroweak parameters, denoted as $S, T, U, V, W, X, Y$, which describe the leading one-loop corrections to the self-energies of electroweak gauge bosons~\cite{Maksymyk:1993zm,Burgess:1993mg,Kundu:1996ah,Barbieri:2004qk}. While $S, T, W,$ and $Y$ are the leading parameters within their respective symmetry classes in the EFT expansion~\cite{Barbieri:2004qk}, their numerical impact varies significantly depending on the mass scale of the new physics.

\begin{figure}[h!]
    \centering
    \includegraphics[width=0.6\linewidth]{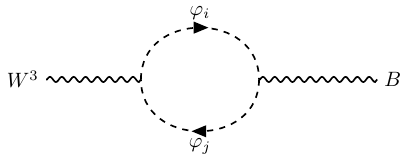}
    \caption{The dominant one-loop vacuum polarization diagram contributing to the oblique parameter $S$. The external neutral gauge bosons, $W^3$ and $B$, mix via a loop of scalar Loryon mass eigenstates $\varphi_i$ and $\varphi_j$. }
    \label{fig:s_param}
\end{figure}

The $T$ parameter vanishes at one-loop order due to the custodial symmetry imposed on the Loryon sector. The $W$ and $Y$ parameters correspond to higher-order terms in the derivative expansion ($\mathcal{O}(p^4)$) compared to the $S$ parameter ($\mathcal{O}(p^2)$). Consequently, for Loryons with masses $m_i > m_W$, the contributions to $W$ and $Y$ are suppressed by a factor of $m_W^2/m_i^2$ relative to $S$, in addition to receiving a numerically smaller scalar form factor contribution~\cite{Banta:2021dek}. We therefore neglect $W$ and $Y$ and focus our attention on the $S$ parameter:
\beq
S = - \frac{4\cos\theta_W \sin\theta_W}{\alpha_{EM}}\, \Pi'_{3B} (p^2=0) \,, 
\eeq
where $\Pi'_{3B}$ represents the derivative of the vacuum polarization amplitude mixing the neutral $SU(2)_L$ boson ($W^3$) and the hypercharge boson ($B$), as shown in figure~\ref{fig:s_param}.

For a scalar Loryon belonging to the custodial representation $[L, R]_X$, the contribution arises from loops of the $n=\dim(L) \times \dim(R)$ mass eigenstates. We denote these mass eigenstates as $\varphi_1, \dots, \varphi_n$ to distinguish them from the $SU(2)_V$ multiplets $\phi_V$. Summing over these physical states, the result is  (from Ref.~\cite{Banta:2021dek})
\beq
\Delta S = \frac{2}{\pi}\, \sum_{i,j=1}^n T_{ij}^3\, Y_{ji}\, \Pi'_S \left( m_i, m_j \right) \,,
\label{eqn:SWYScalar}
\eeq
where the scalar form factor is given by
\beq
\Pi'_S \left( m_i, m_j \right) = \frac{1}{2^{\rho_i}}\,\int_0^1 \dd x\, \Bigg[ x(1-x) \log \frac{\mu^2}{x m_i^2 + (1-x) m_j^2} \Bigg] \,. 
\label{eqn:PiSp}
\eeq
Critically, $\Delta S$ is only non-zero if there is a mass splitting among the Loryon components. 
To quantify this splitting relative to the overall mass scale, we define the dimensionless parameter $r_\text{split}$ in terms of the parameters in \Eq{eq:sl0lag}:
\beq
r_{\text{split}} \equiv \frac{B}{2M^2/v^2 + A}.
\eeq

Given a $2\sigma$ bound of $S \le 0.14$~\cite{ParticleDataGroup:2024cfk}, this can be translated into a constraint for the $[2,2]_0$ Loryon representation that we study. For the other representations that we study, since there is only one representation of $SU(2)_V$, there is no mass gap and no constraint arising from the bound on $S$. For $[2,2]_0$, the bound on $S$ constraints the splitting parameter to the range $r_{\text{split}} \in (-0.67, 1.98)$~\cite{Banta:2021dek}.

\subsection{Higgs Decay Constraints}
The non-decoupling nature of Loryons makes Higgs coupling measurements a sensitive probe of the model. While Loryons interact with the Higgs through renormalizable operators, the approximate $\mathbb{Z}_2$ symmetry assigned to them forbids mixing with the Higgs at tree level. Consequently, Loryons contribute to Higgs couplings only at loop level.
The $h\gamma\gamma$ effective coupling provides the most promising avenue to probe Higgs-Loryon couplings, as this coupling is also abosent in the SM at tree level, and can be sensitively measured in a clean channel at the LHC. 
We follow standard notation and paramerize the deviation of this effective coupling from its SM value by a multiplicative factor $\kappa_\gamma\equiv g_{h\gamma\gamma} / g_{h\gamma\gamma, SM}$.

The tree-level Higgs-Loryon-Loryon coupling for a scalar Loryon $\phi$ reads
\begin{equation}
\mathcal L \supset - \frac{1}{2^\rho} f_\phi \frac{2 m_\phi^2}{v} h\, |\phi|^2  \, ,
\end{equation}
where $f_\phi$ is $f_V$ for the scalar field $\phi$, see \Eq{eq:fvsc}. $\kappa_\gamma$ is then given by
\beq
\kappa_\gamma = 1 + \frac{\sum_\text{BSM} f_i\, Q_i^2\, A_{s_i}(\tau_i)}{\sum_\text{SM} f_i\, Q_i^2\, A_{s_i}(\tau_i)} \,,
\label{eq:kap_gam}
\eeq
where the sum in the denominator accounts for the $W^\pm$ bosons, the top, bottom, charm quarks, and the tau lepton, and only BSM states contribute to the numerator. Contributions from other charged SM particles are negligible due to the smallness of their Yukawa couplings. For each contributing particle $i$, $Q_i$ represents its electromagnetic charge; $\tau_i=4m_i^2/m_h^2$ parameterizes the mass; $s_i$ indicates the spin; and the spin-dependent form factors $A_{s_i}(\tau)$, as defined in Ref.~\cite{Banta:2021dek}, are expressed as
\bealg
\label{eqn:formfactors}
A_0(\tau) &= \tau \big[ 1 - \tau F(\tau) \big] \,, \\
A_{1/2}(\tau) &= -2\tau \big[ 1 + (1-\tau) F(\tau) \big] \,, \\
A_1(\tau) &= 2+3\tau \big[ 1 + (2-\tau) F(\tau) \big] \,,
\eealg
with
\begin{equation}
F(\tau) = \left\{\begin{array}{ll}
\arcsin^2\left( 1/\sqrt{\tau} \right) &\qquad \tau\ge 1 \\[5pt]
-\frac14 \left[ \log\frac{1+\sqrt{1-\tau}}{1-\sqrt{1-\tau}} - i\pi \right]^2 &\qquad \tau<1
\end{array}\right. \,.
\label{eqn:Ftau}
\end{equation}
As the particles in the loop become extremely heavy, specifically in the limit $\tau_i\to\infty$, the form factors asymptote to the values:
\begin{align}
A_0 \to -1/3 \,,\qquad   A_{1/2} \to -4/3 \,,\qquad   A_1 \to 7 \,.
\label{eqn:asymptote}
\end{align}
For particles heavier than the Higgs, which covers most Loryons of interest, this approximation holds well with an error margin of around $\lesssim 10\%$.

To determine the experimental constraints on $\kappa_\gamma$, we rely on the latest results from both the ATLAS and CMS collaborations. Specifically, we use their joint fits to $\kappa_\gamma$, assuming no deviations in tree-level Higgs couplings and no significant contributions from untagged or invisible Higgs decays. Neglecting these small effects, the $2\sigma$ allowed region from ATLAS is $\left| \kappa_\gamma \right| \in (0.89, 1.22)$~\cite{ATLAS:2022vkf}, while CMS reports $\left| \kappa_\gamma \right| \in (0.92, 1.13)$~\cite{CMS:2022dwd}. In order to be conservative, we will demand that $\kappa_{\gamma}$ be consistent with the $2\sigma$ range of both experiments,
\beq
    |\kappa_\gamma| \in (0.92,1.13)\, .
    \label{eq:kap_gam_comb}
\eeq
We will use the full expressions for the $A_{s_i}$ in our calculations, even though for the majority of our parameter region of interest, the asymptotic values of Eq.~\eqref{eqn:asymptote} are a good approximation.

\section{Loryons coupled to a 2HDM}
\label{sec:twoHiggs}

Having reviewed the phenomenology of Loryons for the familiar single Higgs doublet case, we now come to our main focus for this study, where we extend the Higgs sector to a 2HDM. The most general scalar potential of the 2HDM has the form
\bealg
V_{\rm tree} =& m_{11}^2 |H_1|^2 + m_{22}^2 |H_2|^2 - m_{12}^2 (H_1^{\dagger} H_2 + \rm h.c.) \\
&+ \frac{\lambda_1}{2} |H_1|^4 +
\frac{\lambda_2}{2} |H_2|^4 + \lambda_3
|H_1|^2|H_2|^2 + \lambda_4 |H_2^{\dagger} H_1|^2 \\
&+ \left[\frac{\lambda_5}{2} (H_1^{\dagger} H_2)^2 + \lambda_6 |H_1|^2(H_1^{\dagger} H_2)
+\lambda_7 |H_2|^2(H_1^{\dagger} H_2) +\rm
h.c.\right],
\label{eq:V2HDM}
\eealg
where $H_{1,2}$ are the two Higgs doublets. We follow the standard convention to parameterize the VEVs as
\beq
H_{1,2} = 
    \begin{bmatrix}
        \phi_{1,2}^+ \\ \frac{1}{\sqrt2}(v_{1,2} + \phi^0_{1,2} + i\sigma_{1,2})
    \end{bmatrix},
\eeq
with $v_1^2 + v_2^2 = v^2$ and $\tan\beta \equiv v_2 / v_1$. The mass eigenstates are obtained through the mixing of these degrees of freedom. The mixing of the CP-even degrees of freedom is governed by the angle $\alpha$, while the mixing in the charged and CP-odd degrees of freedom is determined by the angle $\beta$:
\begin{eqnarray}
\left(\begin{array}{c}H \\ h \end{array}\right) =  \left(\begin{array}{cc}\cos\alpha & \sin\alpha \\ -\sin\alpha & \cos\alpha \end{array}\right)  \left(\begin{array}{c} \phi^0_1 \\ \phi^0_2 \end{array}\right) , \\
\left(\begin{array}{c}G^0 \\ A \end{array}\right) =  \left(\begin{array}{cc}\cos\beta & \sin\beta \\ -\sin\beta & \cos\beta \end{array}\right)  \left(\begin{array}{c} \sigma_1 \\ \sigma_2 \end{array}\right) , \\
\left(\begin{array}{c}G^{\pm} \\ H^{\pm} \end{array}\right) =  \left(\begin{array}{cc}\cos\beta & \sin\beta \\ -\sin\beta & \cos\beta \end{array}\right)  \left(\begin{array}{c} \phi^{\pm}_1 \\ \phi^{\pm}_2 \end{array}\right) , \,
\end{eqnarray}
where $G^0$ and $G^\pm$ are the Goldstone modes eaten by the $Z$ and $W$ bosons. Note that from this point onwards we use the symbol $H$ (without a subscript) for the heavier CP-even physical Higgs mode, not for a doublet field.

To ensure phenomenological viability, we restrict our analysis to the region of 2HDM parameter space that is consistent with current collider constraints~\cite{ATLAS:2024lyh}.
To prevent tree-level flavor-changing neutral currents (FCNCs) and to simplify the interactions with fermions, an additional discrete $\mathbb{Z}'_2$ symmetry is often imposed on the scalar potential~\cite{Ivanov:2005hg,Davidson:2005cw,Glashow:1976nt}, and we adopt the same choice. Under this symmetry, the scalar fields transform as
\beq
H_1 \to H_1, \qquad H_2 \to -H_2,
\eeq
which sets $m_{12} = \lambda_6 = \lambda_7 = 0$. With this choice, the scalar sector can be fully parametrized by the five physical input parameters:
\beq
m_H, \qquad m_{H^\pm}, \qquad m_A, \qquad \tan\beta, \qquad \cos(\beta - \alpha) . \,
\label{eq:2hdmprm}
\eeq
The relevant triple and quartic couplings from the 2HDM used in this study are listed in Appendix~\ref{app:2hdmfr}.

Current constraints strongly favor the 2HDM alignment limit $\cos(\beta-\alpha) \to 0$, where the light neutral Higgs is SM-like~\cite{ATLAS:2024lyh}. Furthermore, due to the non-decoupling nature of a 2HDM, unitarity constraints disfavor heavy masses for the BSM Higgs states~\cite{Goodsell:2018fex}. With these constraints in mind, we focus our attention on a  benchmark 2HDM spectrum of $(m_H,m_{H^\pm},m_A)=(380,450,450)\ \text{GeV}$. This choice is motivated by two guiding principles: ensuring that the model is consistent with current bounds while maximizing the accessible parameter space for Loryons. For the former, while Type-II and Type-flipped models require a heavier charged Higgs to satisfy flavor constraints, our chosen spectrum remains fully viable for Type-I and Lepton-Specific 2HDM scenarios. For the latter, as discussed in the introduction, increasing the masses of the 2HDM scalars generally constricts the unitarity-allowed window for Loryons. By selecting 2HDM masses that are as light as experimentally permitted, we thereby maximize the potential Loryon parameter space. 

For this benchmark spectrum, we consider several benchmark values for $\tan\beta$ and $\cos(\beta - \alpha)$, in order to sample the viable region in the 2HDM parameter space. We include two benchmark points at the alignment limit $\cos(\beta - \alpha)=0$ with $\tan\beta=1$ and $\tan\beta=1.7$. We also include two additional benchmark points to consider small deviations from the alignment limit, with $\cos(\beta-\alpha)=-0.1$ and $\cos(\beta-\alpha)=0.02$, and $\tan\beta=1.5$ for both. The results of our analysis will be presented for these points in the 2HDM parameter space.

With the approximate $\mathbb{Z}_2$ symmetry for the Loryons and the additional $\mathbb{Z}'_2$ symmetry on the Higgs bi-doublets $\mathcal H_{1,2}$, the most general set of interactions becomes~\eqref{eq:sl0lag}:
\beq
\mathcal L \supset -\frac{M^2}{2^\rho} \, \tr(\Phi^\dagger \Phi) 
- \frac{A_{ij}}{2^\rho}\, \tr(\Phi^\dagger \Phi)\, \ha\, \tr(\mathcal H^\dagger_i \mathcal H_j) 
- \frac{B_{ij}}{2^\rho}\, 2\,\tr(\Phi^\dagger T_{L}^a \Phi T^{\dot a}_{R})\,2\,\tr(\mathcal H_i^\dagger T_{2}^a \mathcal H_j T_{2}^{\dot a}),
\label{eq:sl1lag}
\eeq
where a sum over $i,j$, indexing the two Higgs doublets, is implied. The two discrete symmetries eliminate the off-diagonal components of the coupling matrices $A_{ij}$ and $B_{ij}$, thereby reducing the number of independent couplings to four per Loryon representation\footnote{Of course, if there is more than one Loryon in a given representation, then $M^2$, $A$ and $B$ also become matrices in Loryon flavor space.}. As with a single Higgs doublet, the $B_{ij}$ term is only included in the Lagrangian when $L$ and $R$ are both non-singlet representations.

Consequently, after EWSB, Eq.\eqref{eq:sl0lagew} for the 2HDM generalizes to:
\beq
\mathcal L \supset -\frac{1}{2^\rho} \sum_{V\in\mathcal V}\phi_V^\dagger \left\{ M^2 + \frac12 \lambda_{Vij} (v_i + h_i)(v_j + h_j) \right\} \phi_V,
\label{eq:slm2hdm}
\eeq
where
\beq
\lambda_{Vij} = A_{ij} + B_{ij}[C_2(L)+C_2(R) - C_2(V)].
\eeq
As before, $L$, $R$, and $V$ denote the representations under $SU(2)_L$, $SU(2)_R$, and the diagonal subgroup $SU(2)_V$, respectively. The quantity $C_2$ is the quadratic Casimir invariant, as defined in Eq.~\eqref{eq:C2}.
Note that the couplings to $H^\pm$ and $A$ are irrelevant to the constraints on Loryons that we consider, and they will not appear in the rest of our analysis.

\subsection{HEFT Condition}
Determining the conditions under which HEFT \footnote{Note that because both Higgs doublets acquire vacuum expectation values, the effective field theory expanded around this vacuum could more precisely be termed 2HDMEFT. However, to avoid introducing additional terminology, we simply refer to it as HEFT throughout this work.} in the 2HDM is considerably more challenging. However, in the regime where the mass of the Loryon is larger than that of all Higgs states, one can explicitly demonstrate that the criterion for HEFT being the only valid low-energy effective description -- i.e., when SMEFT ceases to provide an adequate approximation -- remains unchanged: more than half of the heaviest scalar Loryon’s mass must originate from the Higgs vacuum expectation value, see \Eq{eq:fmax}. This result follows from integrating out the Loryons at one loop and studying the analytic structure of the resulting effective Lagrangian, as detailed in Appendix~\ref{app:heft}.
We find that HEFT is the only valid description for describing the Loryons when the following set of conditions hold:
\begin{equation}
    \frac{\ha\sum_{ij}[A_{ij}+\b{C_2(L)+C_2(R)-C_2(V)}B_{ij}]v_i v_j}{m_V^2}>\frac{1}{2} \ \ \  \forall V\in\mathcal{V}.
\end{equation}
Although this conclusion is derived in the heavy-Loryon regime, we adopt it as a working definition of HEFT-relevant Loryons throughout this paper: any scalar state, which receives more than half of the contribution to its mass squared from electroweak symmetry breaking is a Loryon.

\subsection{Electroweak Precision Measurements}
The extension of the scalar sector to a 2HDM potentially alters the constraints imposed by electroweak precision observables. However, at the one-loop level, the contributions to the oblique parameters $S$ and $T$ from the Loryon sector and the 2HDM scalar sector are additive and independent. 

Consequently, the contribution of a scalar Loryon to the $S$ parameter remains identical in functional form to the single Higgs doublet case presented in Section~\ref{ssec:ewpm}. The total theoretical prediction is given by
\begin{equation*}
S_{\text{total}} = S_{\text{SM}} + S_{\text{2HDM}} + S_{\text{Loryon}}.
\end{equation*}
Experimental constraints apply to the deviation from the Standard Model, $\Delta S = S_{\text{2HDM}} + S_{\text{Loryon}}$. The 2HDM contribution, $S_{\text{2HDM}}$, depends on the mass splitting between the heavy neutral and charged Higgs bosons ($H, A, H^{\pm}$) and the mixing angles. 

Compared to the case with a single Higgs doublet, a non-zero contribution from the 2HDM sector shifts the central value of the allowed window for $S_{\text{Loryon}}$, potentially tightening or relaxing the constraints on the Loryon mass splittings. Since in order to stay with experimental constraints we only consider benchmark points at or near the alignment limit, and the BSM scalars are heavy, $S_{\text{2HDM}}$ remains small.

We remind the reader that the Loryon contribution to the $S$-parameter is only nonzero for the $[2,2]_0$ representation that we consider, due to there being a mass splitting between the singlet and triplet components. Even in this case, the contribution to the $S$-parameter, given by \Eq{eqn:SWYScalar} is within the experimentally allowed range, and places no constraints on the parameter space of interest.

Next, we turn to the phenomenology of specific Loryon representations. The introduction of a second Higgs doublet significantly increases the complexity of the scalar potential and the resulting scattering matrices. Consequently, we restrict our detailed analysis to a selected subset of $[L,R]_X$ representations that remain phenomenologically viable and distinct: the neutral and charged singlets $[1,1]_{0,1}$, the triplet $[1,3]_0$ (and the equivalent $[3,1]_0$), and the bi-doublet $[2,2]_0$.

\subsection{$[1,1]_{0}$ and $[1,1]_{1}$ Representations}
\label{sec:LR11}
For pedagogical clarity, we begin with the simplest representation, $[1,1]_X$, before turning to more elaborate cases. In this setting, the relevant terms in Lagrangian for the singlet Loryon field $\phi$ takes the form
\begin{multline}
\ML \supset 
-\frac{1}{2 \cdot 12^\rho}\lambda_{4}|\phi|^{4}
-\frac{1}{2^\rho} M^{2}|\phi|^{2} \\
-\frac{1}{2\cdot2^\rho}A_{11}|\phi|^{2}\b{v \cos\beta - h\sin\alpha + H\cos\alpha}^{2} \\
-\frac{1}{2\cdot2^\rho}A_{22}|\phi|^{2}\b{v \sin\beta + h\cos\alpha + H\sin\alpha}^{2},
\end{multline}
where $\rho = 1$ for $X = 0$ and $\rho = 0$ for $X \ne 0$. Note that when $X \neq 0$, $\phi$ carries electric charge and couples to photons, thereby inducing additional constraints from Higgs decay channels which we study below.

While the couplings $A_{11}$ and $A_{22}$ define the interactions with the Higgs basis, the phenomenology, including the mass spectrum and unitarity bounds, is most conveniently described by the linear combinations:
\beq
C_{\pm} \equiv A_{11} \pm A_{22}.
\label{eq:cpm11}
\eeq
Substituting these into the Lagrangian, the physical mass of $\phi$ is given by
\begin{equation}
m^2 = M^2 + \frac{v^2}{4}(C_+ + C_- \cos 2\beta).
\label{eq:LR11Ymass}
\end{equation}

We next turn to the phenomenological implications of this representation, in particular the constraints from perturbative unitarity and from Higgs decay channels in the presence of an additional Higgs doublet.

\subsubsection{Unitarity Bounds}
Upon including the neutral heavy Higgs field $H$ as a possible initial/final state, the zeroth partial-wave coefficient matrix in Eq.~\eqref{eq:a0m1} enlarges from its original $[2\times2]\oplus [1]$ structure to a $[4\times4] \oplus [2\times2]$ form, given by
\bealg
\, 
\begin{bmatrix}
    a_0^{\phi \phi \to \phi \phi} &  a_0^{\phi \phi \to hh} & a_0^{\phi \phi \to hH} &  a_0^{\phi \phi \to HH}\\
    a_0^{hh \to \phi \phi} & a_0^{hh \to hh}  &  a_0^{hh \to hH} & a_0^{hh \to HH} \\
    a_0^{hH \to \phi \phi} & a_0^{hH \to hh}  &  a_0^{hH \to hH} & a_0^{hH \to HH} \\
    a_0^{HH \to \phi \phi} & a_0^{HH \to hh}  &  a_0^{HH \to hH}  & a_0^{HH \to HH} \\
\end{bmatrix} \oplus
\begin{bmatrix}
    a_0^{\phi h \to \phi h} &  a_0^{\phi h \to \phi H}\\
    a_0^{\phi H \to \phi h} & a_0^{\phi H \to \phi H}
\end{bmatrix} 
\eealg
in the basis $\{ \phi \phi, h h , h H , H H, \phi h, \phi H \}$.
The amplitudes are primarily controlled by the trilinear couplings involving the Loryon and the physical Higgs states
\beq
\ML \supset -\ha \, v \, c_h |\phi|^2 h - \ha \, v \, c_H |\phi|^2 H,
\label{eq:chcH11}
\eeq
where the coupling coefficients $c_h$ and $c_H$ are expressed in terms of the $C_{\pm}$ parameters defined previously
\begin{align}
c_{h} & =\frac{1}{2^\rho}\s{C_{+}\sin\b{\beta-\alpha}-C_{-}\sin\b{\alpha+\beta}}\\
c_{H} & =\frac{1}{2^\rho}\s{C_{+}\cos\b{\beta-\alpha}+C_{-}\cos\b{\alpha+\beta}}.
\end{align}
As discussed in Section~\ref{sec:ub0}, quartic contributions such as $|\phi|^4$, $|\phi|^2 h^2$, and $|\phi|^2 H^2$ never drive the unitarity bounds and will not appear further in this analysis. Note that we only include the $CP$-even Higgs states $h$ and $H$ in the basis of initial and final states for scattering. This is to reduce the dimensionality of the $[a_0]$ matrix. Since the remaining Higgs states are related to these via transformations of the custodial group, their omission does not introduce a loss of generality.

\subsubsection{Higgs Decay Constraints}
Constraints from Higgs decays, particularly the diphoton channel $h \to \gamma\gamma$, impose restrictions on any representations containing states of non-zero electric charge. While the neutral $[1, 1]_0$ representation lacks electromagnetic interactions and thus evades these bounds, the $[1, 1]_1$ case contains a charged scalar that contributes to the decay rate at the one-loop level.

The experimental constraint is derived from the combined ATLAS and CMS measurements~\cite{ATLAS:2024lyh,CMS:2022dwd}, and as described in the previous section, we demand $|\kappa_{\gamma}| \in (0.92, 1.13)$. In the 2HDM framework, the total BSM contribution to $\kappa_\gamma$ arises from the sum of the charged Loryon loop and the 2HDM charged Higgs ($H^\pm$) loop. The effective BSM term in Eq.~\eqref{eq:kap_gam} becomes
\beq
\sum_\text{BSM} f_i\, Q_i^2\, A_{s_i}(\tau_i) = \frac{v^2 \, c_h}{2m_\phi^2} A_0 \left( \tau_{\phi} \right) + \frac{v g_{hH^{+} H^{-}}}{2m_{H^\pm}^2} A_0 \left( \tau_{H^\pm} \right),
\label{eq:kap_gam2}
\eeq
where $c_h$ is the Loryon coupling to the light Higgs defined in Eq.~\eqref{eq:chcH11} and $g_{h H^{+} H^{-}}$ is the 2HDM trilinear coupling.

For our analysis, we adopt a benchmark charged Higgs mass of $m_{H^\pm} = 450 \gev$. Applying the experimental limits, we obtain the following numerical bound on the Loryon parameter space
\beq
\begin{aligned}
    &\frac{v^2 c_h}{2m_\phi^2}A_0(\tau_\phi)Q^2+\frac{v g_{hH^{+} H^{-}}}{2m_{H^\pm}^2}A_0(\tau_{H^\pm})\in (-13.97,-12.6)\cup (-0.52,0.85).
     \label{eq:higgsdecay11}
\end{aligned}
\eeq
Because the experimental bounds apply to the absolute value $|\kappa_\gamma|$, the allowed parameter space splits into two regions corresponding to $\kappa_{\gamma} \in (-1.13,-0.92)$ and $\kappa_{\gamma} \in (0.92, 1.13)$, respectively. This inequality explicitly demonstrates how the allowed region for the Loryon coupling $c_h$ is shifted by the 2HDM parameters $\alpha$ and $\beta$ through the charged Higgs contribution. We now combine these constraints with those from unitarity, and plot the allowed region in parameter space.

\subsubsection{Results}
\begin{figure}[!h]
  \centering

  \begin{subfigure}[b]{0.49\textwidth}
    \includegraphics[width=\textwidth]{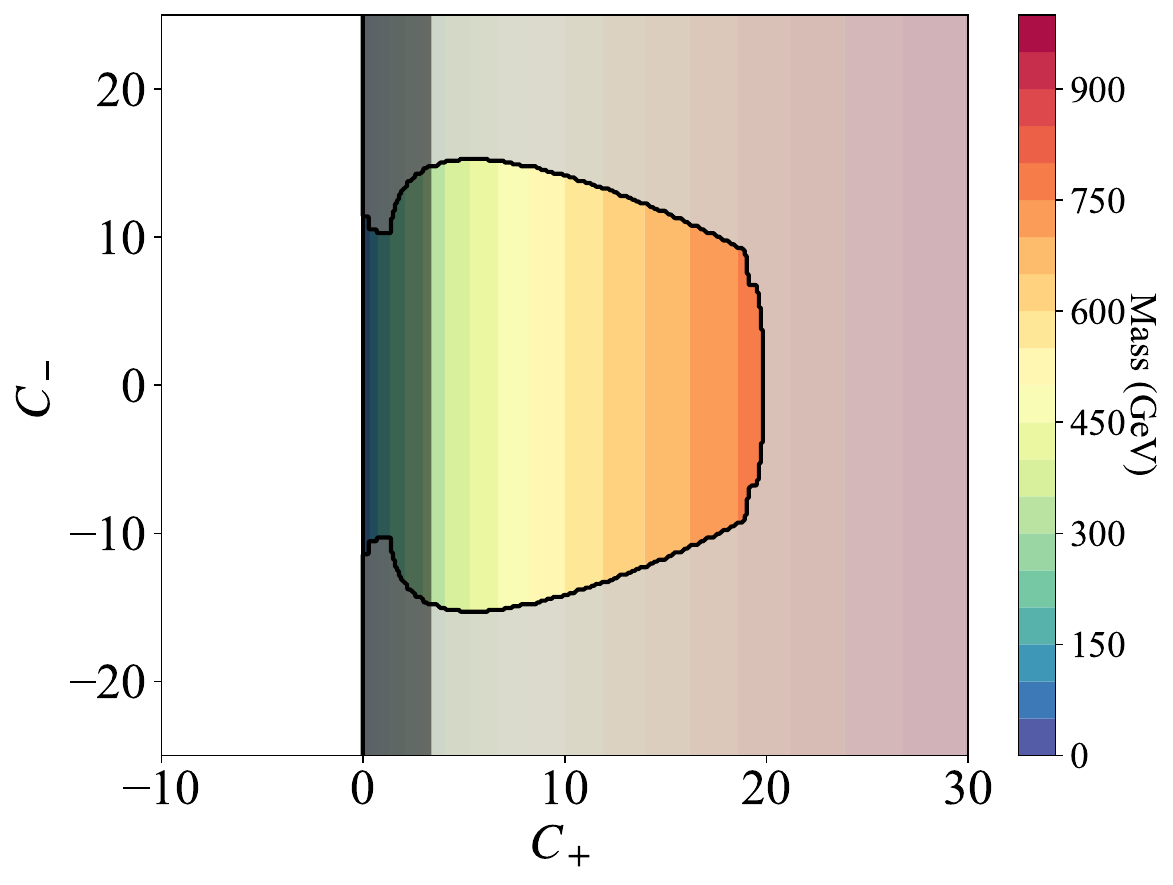}
    \caption{$(0,1)$}
  \end{subfigure}
  \hfill
  \begin{subfigure}[b]{0.49\textwidth}
    \includegraphics[width=\textwidth]{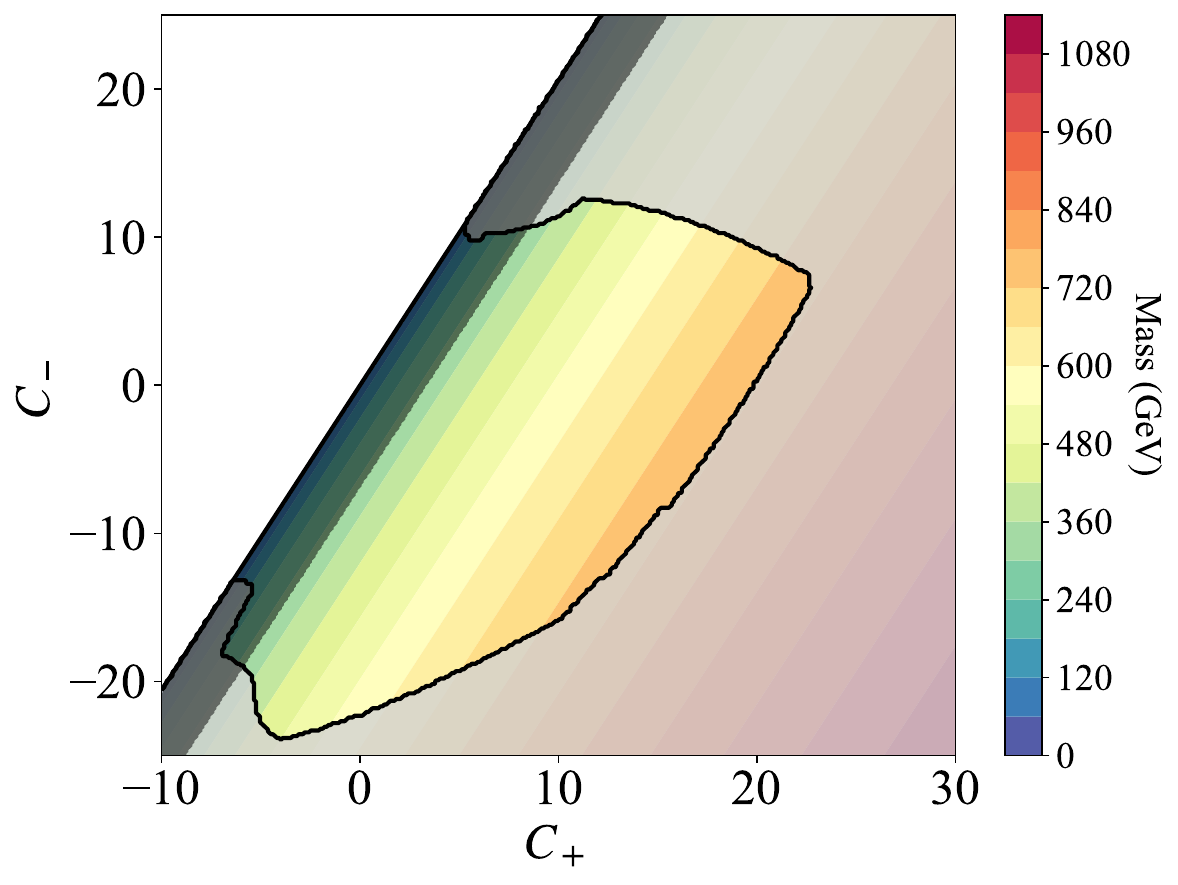}
    \caption{$(0,1.7)$}
  \end{subfigure}
  
  \begin{subfigure}[b]{0.49\textwidth}
    \includegraphics[width=\textwidth]{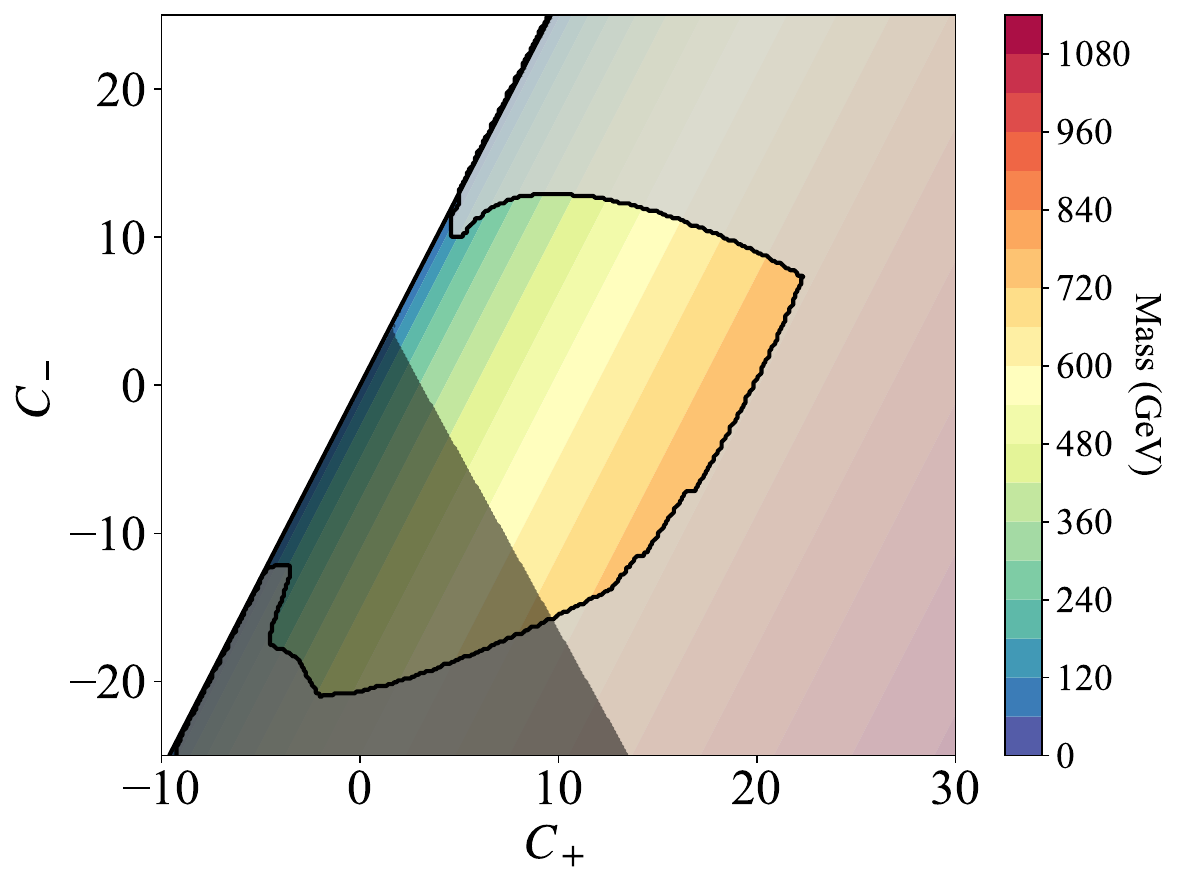}
    \caption{$(0.02,1.5)$}
  \end{subfigure}
  \hfill
  \begin{subfigure}[b]{0.49\textwidth}
    \includegraphics[width=\textwidth]{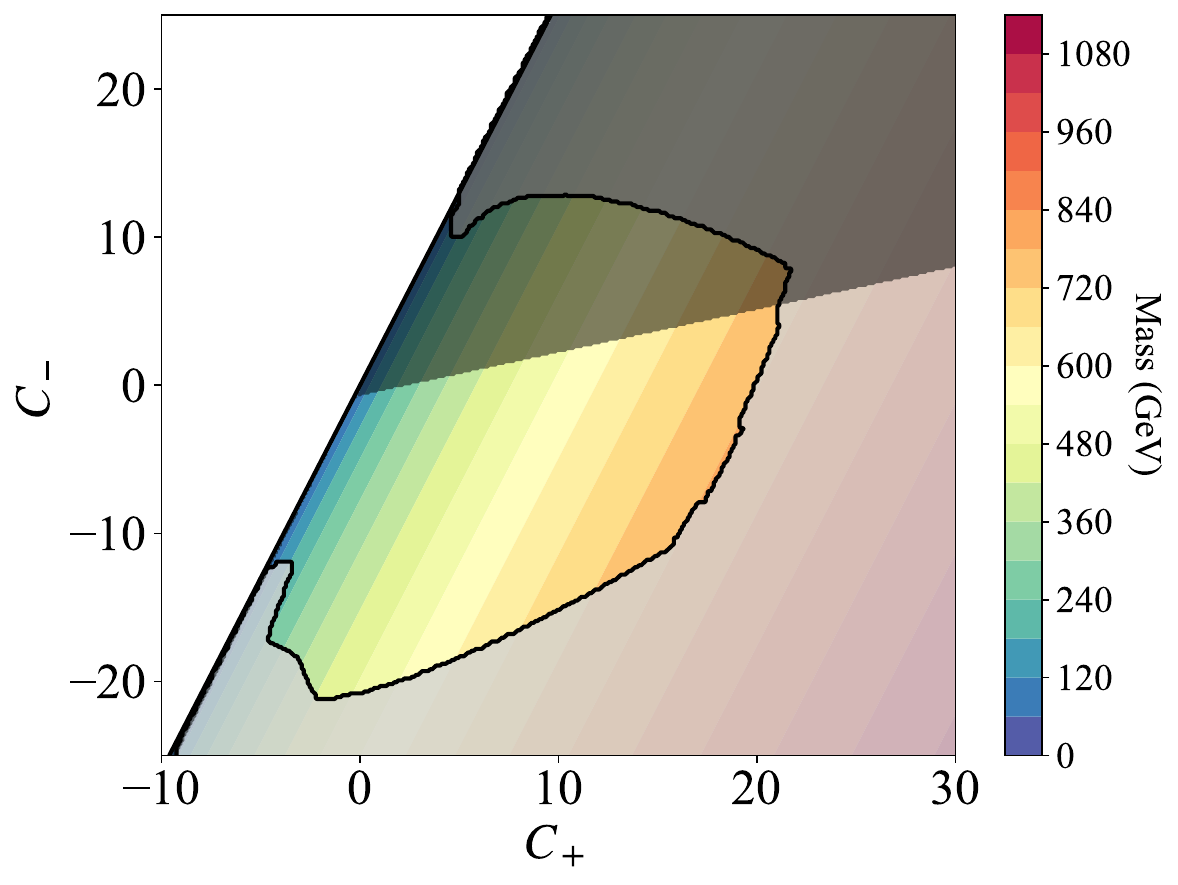}
    \caption{$(-0.1,1.5)$}
  \end{subfigure}
  \caption{The allowed parameter space for the $[1,1]_X$ Loryon, taking into account constraints from perturbative unitarity and Higgs decay measurements, for fixed fraction $f_\phi=0.5$ and different values of $(\cos(\beta-\alpha),\tan\beta)$ as indicated at the bottom of each panel. The part of the parameter space outside the colored region is ruled out by unitarity bounds and the dark-shaded region is ruled out by the $\kappa_\gamma$ bound. 
  The white region is unphysical as ($m_\phi^2 < 0$) there. Inside the allowed region, the color gradient from blue to red represents the physical mass of the Loryon.}
  \label{fig:LR11Yf05}
\end{figure}

\begin{figure}[!h]
  \centering

  \begin{subfigure}[b]{0.49\textwidth}
    \includegraphics[width=\textwidth]{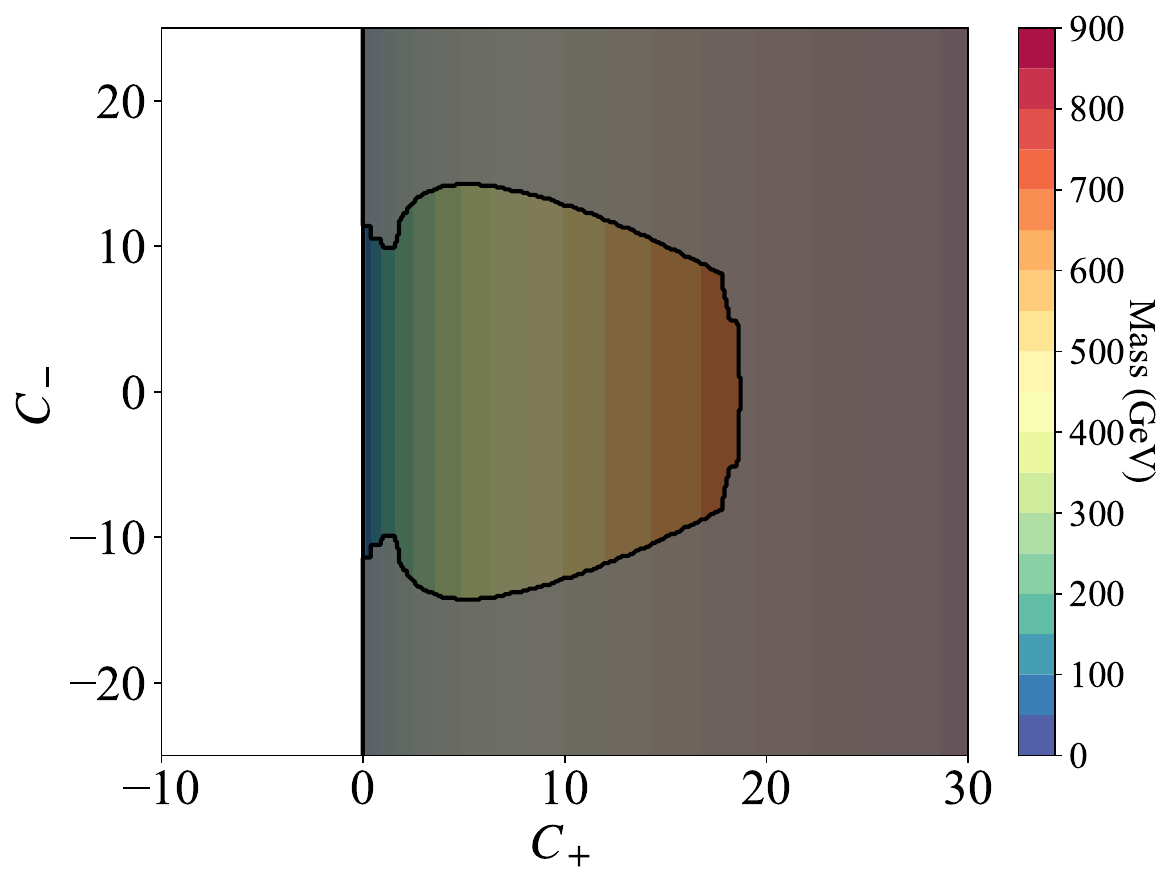}
    \caption{$(0,1)$}
  \end{subfigure}
  \hfill
  \begin{subfigure}[b]{0.49\textwidth}
    \includegraphics[width=\textwidth]{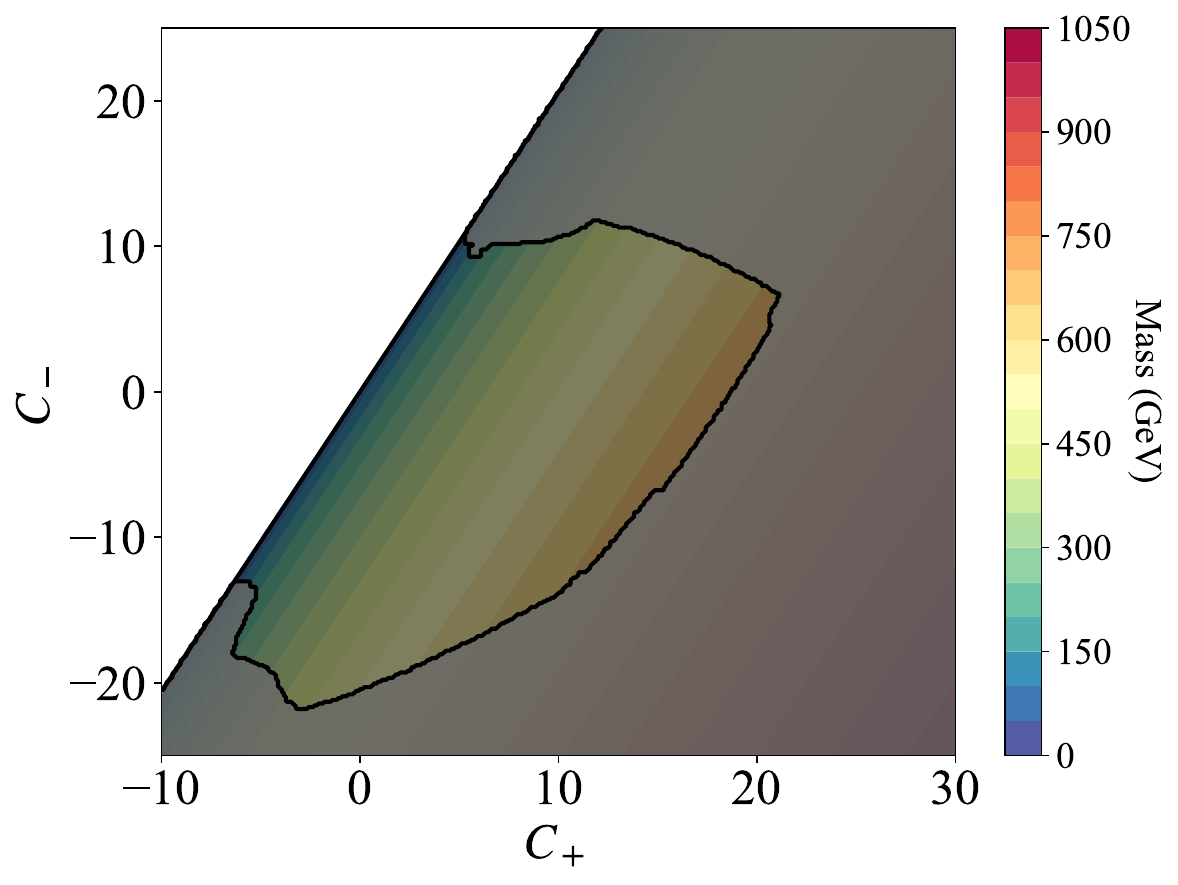}
    \caption{$(0,1.7)$}
  \end{subfigure}
  \begin{subfigure}[b]{0.49\textwidth}
    \includegraphics[width=\textwidth]{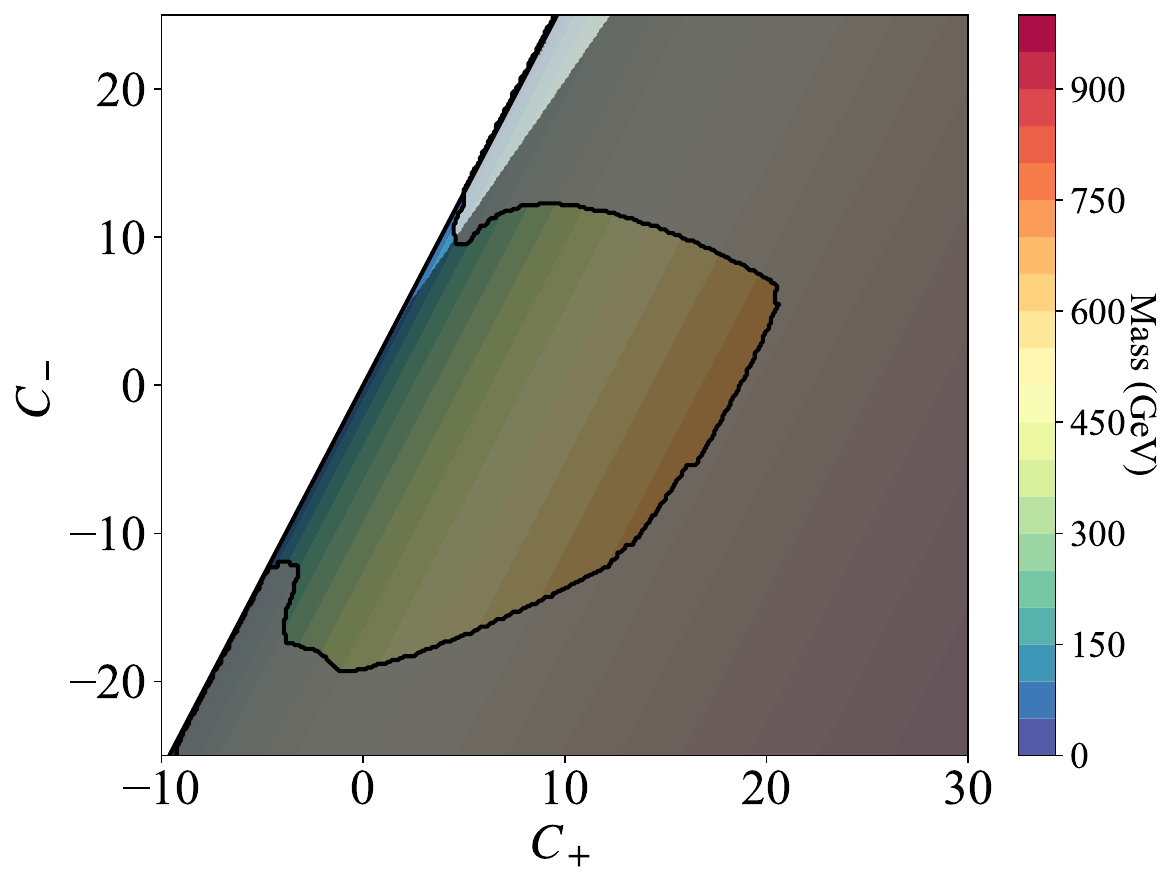}
    \caption{$(0.02,1.5)$}
  \end{subfigure}
  \hfill
  \begin{subfigure}[b]{0.49\textwidth}
    \includegraphics[width=\textwidth]{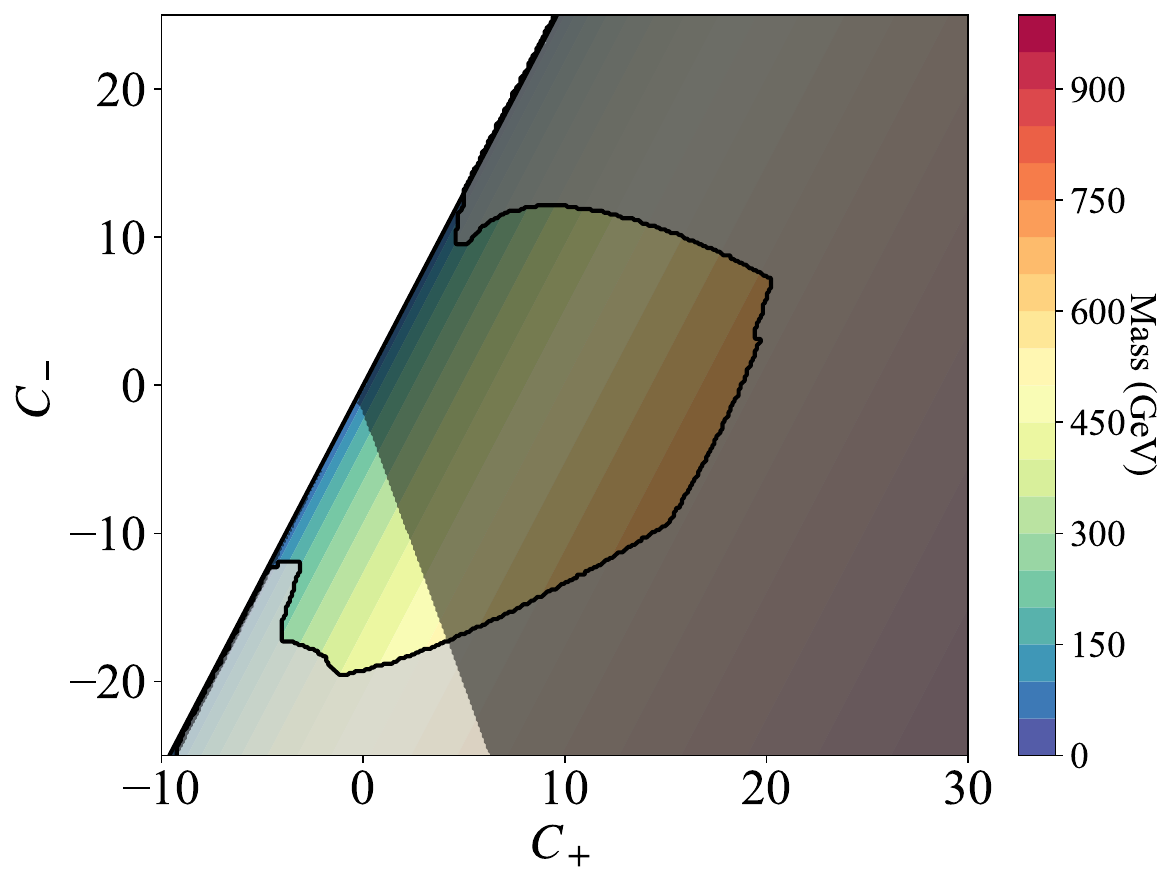}
    \caption{$(-0.1,1.5)$}
  \end{subfigure}
  \caption{Similar to Figure~\ref{fig:LR11Yf05}, but with the mass fraction fixed to $f_{\phi}=0.6$. 
  }
  \label{fig:LR11Yf0.6}
\end{figure}

We present the combined constraints on the $[1, 1]_X$ representation in Figures~\ref{fig:LR11Yf05} and \ref{fig:LR11Yf0.6}, projected onto the $(C_+, C_-)$ plane. As defined in Eq.~\eqref{eq:cpm11}, the axes correspond to the symmetric and antisymmetric combinations of the couplings to the two Higgs doublets. A fundamental stability requirement on $C_+$ is that the Loryon mass squared remains positive given $\alpha$ and $\beta$.

In these plots, we fix the mass fraction originating from EWSB to $f_\phi = 0.5$ (Figures~\ref{fig:LR11Yf05}) and $f_\phi = 0.6$ (Figures~\ref{fig:LR11Yf0.6}). The parameter space is constrained by two primary factors:
\begin{itemize}
    \item {\bf Unitarity (Gray)}: The region shaded in gray is excluded by perturbative unitarity, which places an upper limit on the magnitude of the couplings.
    \item {\bf Higgs Decays (Black)}: The region shaded in black corresponds to constraints from the Higgs diphoton decay rate ($\kappa_\gamma$) for a benchmark of $X=1$. This exclusion region would expand for larger values of $X$.
\end{itemize}
The remaining valid parameter space is colored with a gradient ranging from blue to red, indicating the physical mass of the Loryon corresponding to the value of $C_{\pm}$ at each point and the values of $f_\phi$, $\tan\beta$ and $\cos(\beta - \alpha)$ for each figure. The panels in Figures~\ref{fig:LR11Yf05} and \ref{fig:LR11Yf0.6} illustrate how this viable region shifts as the 2HDM mixing angles $\alpha$ and $\beta$ vary. For the neutral representation $[1, 1]_0$, there is no constraint from Higgs decay, leaving the allowed space determined solely by the unitarity bounds and the mass positivity condition.

The qualitative behavior of the allowed regions follows from the mass relation in Eq.~\eqref{eq:LR11Ymass}. Since the Loryon mass increases with the couplings $C_\pm$ for a fixed EWSB mass fraction $f_\phi$, maintaining a constant $f_\phi$ requires the couplings to grow with the physical mass. Consequently, as the mass increases, the couplings eventually violate perturbative unitarity. This effect becomes more pronounced as the fraction $f_\phi$ increases; by comparing Figure~\ref{fig:LR11Yf0.6} ($f_\phi=0.6$) with Figure~\ref{fig:LR11Yf05} ($f_\phi=0.5$), we observe that a higher EWSB mass fraction significantly tightens the unitarity constraints, shrinking the available parameter space.

In addition, the boundary of the region ruled out by unitarity  exhibits sharp kinks rather than smooth curves. These features arise because at different points in the parameter space, the maximum eigenvalue of the $[a_0]$ matrix is obtained for different scattering channels and different values of $\sqrt{s}$. This is also discussed in Ref.~\cite{Goodsell:2018tti}.

Meanwhile, the $\kappa_\gamma$ bounds depend sensitively on both $\alpha$ and $\beta$, as these parameters govern the coupling of the Loryon to the light Higgs boson. As the angle $\beta$ changes (evident in the last two panels), the specific linear combination of $C_\pm$ contributing to the mass changes, rotating the orientation of the physical mass contours relative to the exclusion bounds.

\subsection{$[1,3]_0$ and $[3,1]_0$ Representations}
We now turn to the next tier of complexity: the representations $[1, 3]_0$ and $[3, 1]_0$. Despite the higher dimensional representation, the scalar potential introduces no new parameters compared to the $[1, 1]_0$ case discussed in the previous section.

We can treat the three scalar degrees of freedom in these representations as a triplet vector $\phi^a$ (where $a=1, 2, 3$) under the $SO(3)$ group associated with the diagonal $SU(2)_V$ group. The relevant interactions with the physical Higgs states take the same form as Eq.~\eqref{eq:chcH11}, summed over the triplet components:
\beq
\ML \supset -\ha \, v \, c_h \phi^a\phi^a h - \ha \, v \, c_H \phi^a\phi^a H,
\label{eq:chcH31}
\eeq
where the coupling coefficients $c_h$ and $c_H$ are defined exactly as in the $[1, 1]$ case.

These states decompose under the SM gauge group $SU(2)_L \times U(1)_Y$ as follows:
\begin{itemize}
    \item {\bf $[1, 3]_0$ Representation}: This transforms as an $SU(2)_L$ singlet. However, unlike the simple neutral singlet in Section~\ref{sec:LR11}, this representation splits into three states with distinct hypercharges $Y = (-1, 0, +1)$. Consequently, this multiplet yields one neutral scalar (real) and an electrically charged scalar (complex).
    \item {\bf $[3, 1]_0$ Representation}: This transforms as a standard $SU(2)_L$ triplet with zero hypercharge ($Y=0$). This multiplet also contains one neutral scalar and one electrically charged scalar.
\end{itemize} 
For a physical Loryon potential the coefficient of the quartic self-coupling is positive; moreover, we are explicitly considering Loryons that do not acquire a vacuum expectation value, which inherently requires $m_V^2 > 0$. As a result, the structure of the scalar potential is identical to the $[1,1]$ case. 

Compared to the $[1,1]_{0,1}$ representations, this case includes a larger number of Loryon states, which raises the dimensionality of the $[a_0]$ matrix considerably. However, different Loryon states are related to each other via $SU(2)_L \times SU(2)_R$ transformations, and it is therefore sufficient to only consider a subset of processes such that all remaining processes are equivalent to them via these transformations. A similar logic can be applied to the Higgs degrees of freedom such as $A$ or $H^{\pm}$. Therefore, to simplify the analysis, we limit ourselves to only considering neutral external states when calculating the $[a_0]$ matrix. With this restriction, the calculation of the unitarity bounds becomes identical to the $[1,1]_{0,1}$ representations, shown in Figures~\ref{fig:LR11Yf05} and~\ref{fig:LR11Yf0.6}. There being one charged scalar in the $[1,3]_0$ and $[3,1]_0$ representations, the Higgs decay bounds are also identical to those of the $[1,1]_{1}$ representation, also shown in the same figures.

\subsection{$[2,2]_0$ Representation}
We now turn to the $[2,2]_0$ representation, which exhibits the richest phenomenology among the cases considered in this work. Following electroweak symmetry breaking, the $[2, 2]$ representation decomposes into two irreducible multiplets under the unbroken custodial $SU(2)_V$: a singlet $s_0$ and a triplet $s_1$, which is treated as a triplet vector.

The relevant interactions with the physical Higgs bosons are described by the following effective Lagrangian
\bealg
\mathcal{L} \supset & 
     \, -\frac{1}{2} v c^0_h \, s_0^2 \, h - \frac{1}{2} v c^0_H \, s_0^2 \,H
     - \frac{1}{2} v c^1_h  \, s_1^a s_1^a \,h - \frac{1}{2}v c^1_H \, s_1^a s_1^a \,H,\, 
\eealg
where the trilinear couplings $c_{h/H}^{i}$ (for $i=0,1$) are linear combinations of the fundamental parameters:
\bealg
    c^i_h &\equiv  \frac{1}{2}\b{C_{+}^i  \sin(\beta-\alpha) + C_{-}^i  \sin(\beta+\alpha) } , \,  \\
    c^i_H &\equiv  \frac{1}{2}\b{C_{+}^i  \cos(\beta-\alpha) + C_{-}^i \cos(\beta+\alpha)}. \,
    \label{eq:chcH22}
\eealg
Here, the parameters $C_{\pm}^{i}$ represent the symmetric and antisymmetric combinations of the couplings to the two Higgs doublets for the singlet ($i=0$) and triplet ($i=1$) states:
\bealg
C_{\pm}^0 = \b{A_{11} + \frac{3}{2}B_{11}} \pm \b{A_{22} + \frac{3}{2}B_{22}} , \, \\
C_{\pm}^1 = \b{A_{11} + \frac{1}{2}B_{11}} \pm \b{A_{22} + \frac{1}{2}B_{22}} . \, \\
\eealg
Substituting these into the scalar potential, the physical masses are determined by the sum of the bare mass $M^2$ and the EWSB contribution:
\begin{equation*}
\begin{aligned}
m_0^2 &= M^2 + \frac{v^2}{4}\b{C_+^0 + C_-^0 \cos 2\beta} \\
m_1^2 &= M^2 + \frac{v^2}{4}\b{C_+^1 + C_-^1 \cos 2\beta}.
\end{aligned}
\end{equation*}

\subsubsection{Unitary Bounds}
With the inclusion of the heavy CP-even Higgs field $H$ and the two distinct Loryon multiplets $s_0$ (singlet) and $s_1$ (triplet) arising from the $[2,2]$ representation, the scattering matrix becomes significantly more complex. The zeroth partial-wave coefficient matrix $[a_0]$, previously defined in Eq.~\eqref{eq:a0m1}, expands to the larger block-diagonal form
\beq
[a_0] \to [5\times5] \oplus [2\times2] \oplus[2\times2] \oplus [1].
\eeq
The dominant $5\times5$ block governs the scattering among the neutral two-particle states. In the basis $\{s_0 s_0, s_1 s_1, hh, hH, HH\}$, this matrix takes the form
\beq
[a_0^{(5)} ]= 
\begin{bmatrix}
    a_0^{s_0 s_0 \to s_0 s_0} & a_0^{s_0 s_0 \to s_1 s_1} &  a_0^{s_0 s_0 \to hh} & a_0^{s_0 s_0 \to hH} &  a_0^{s_0 s_0 \to HH}\\
    a_0^{s_1 s_1 \to s_0 s_0} & a_0^{s_1 s_1 \to s_1 s_1} &  a_0^{s_1 s_1 \to hh}  & a_0^{s_1 s_1 \to hH} & a_0^{s_1 s_1 \to HH} \\
    a_0^{hh \to s_0 s_0} & a_0^{hh \to s_1 s_1} &  a_0^{hh \to hh}  &  a_0^{hh \to hH} & a_0^{hh \to HH} \\
    a_0^{hH \to s_0 s_0} & a_0^{hH \to s_1 s_1} &  a_0^{hH \to hh}  &  a_0^{hH \to hH} & a_0^{hH \to HH} \\
    a_0^{HH \to s_0 s_0} & a_0^{HH \to s_1 s_1} &  a_0^{HH \to hh}  &  a_0^{HH \to hH}  & a_0^{HH \to HH} \\
\end{bmatrix} .
\eeq
Additionally, there are channels involving single Loryon states, specifically the $2\times2$ blocks in the bases $\{s_0 h, s_0 H\}$ and $\{s_1 h, s_1 H\}$, as well as a $[1\times 1]$ channel for $s_0 s_1$ scattering. Note that we are only considering neutral fields as our external states, and therefore every instance of $s_1$ in the matrix above is the neutral component of the triplet $s_1^a$.
Similar to the previous cases, the unitarity bounds are primarily driven by the magnitude of the trilinear couplings $c_{h/H}^i$ defined in Eq.~\eqref{eq:chcH22}, which grow with the Loryon masses.

\subsubsection{Higgs Decay Constraints}
In the $[2,2]$ representation, the constraints from the Higgs diphoton decay $h \to \gamma\gamma$ arise from the charged components within the triplet multiplet $s_1$. The singlet $s_0$ is electrically neutral and does not contribute to this loop-induced process.

The total BSM contribution to the signal strength modifier $\kappa_\gamma$, as introduced in Eq.~\eqref{eq:kap_gam}, is determined by the sum of the charged Loryon loop (from $s_1$) and the 2HDM charged Higgs loop. The effective BSM term reduces to
\beq
\sum_{\text{BSM}} f_i Q_i^2 A_{s_i}(\tau_i) = \frac{v^2 c_h^1}{2 m_1^2} A_0(\tau_{s_1}) Q_{s_1}^2 + \frac{v g_{h H^{+} H^{-}}}{2 m_{H^\pm}^2} A_0(\tau_{H^\pm}),
\eeq
where $c_h^1$ is the coupling of the triplet Loryon to the light Higgs defined in Eq.~\eqref{eq:chcH22}, and $m_1$ is its physical mass. 
The term $g_{h H^{+} H^{-}}$ represents the standard 2HDM trilinear coupling between the light Higgs and the charged Higgs pair again.

As with the $[1,1]$ case, we adopt a benchmark charged Higgs mass of $m_{H^\pm} = 450 \gev$ and require the signal strength to fall within the experimental bounds $\kappa_{\gamma} \in (0.92, 1.13)$. The resulting constraints restrict the parameter space of the triplet coupling $c_h^1$ relative to the 2HDM mixing angles.

\subsubsection{Results}
\begin{figure}[h!]
  \centering
  \begin{subfigure}[b]{0.49\textwidth}
    \centering
    \includegraphics[width=\textwidth]{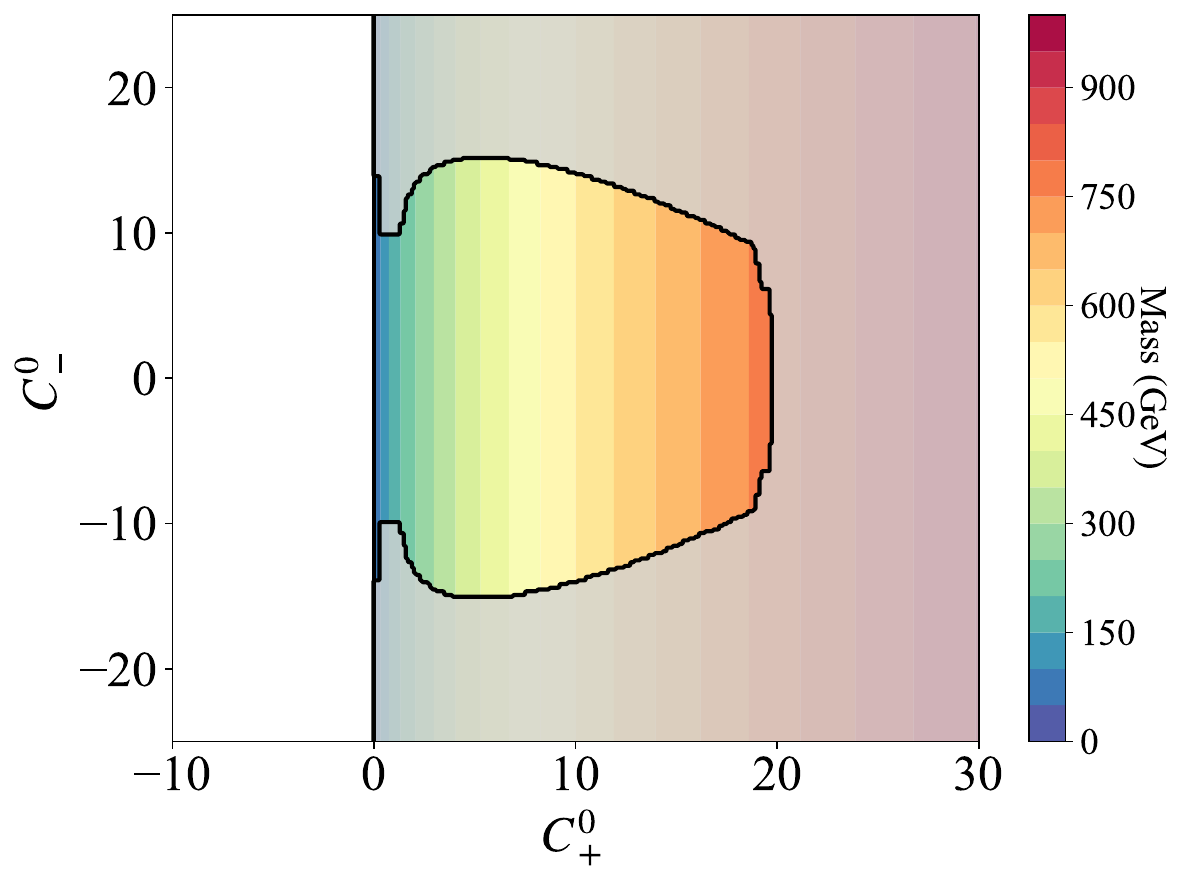} 
  \end{subfigure}
  \hfill
  \begin{subfigure}[b]{0.49\textwidth}
    \centering
    \includegraphics[width=\textwidth]{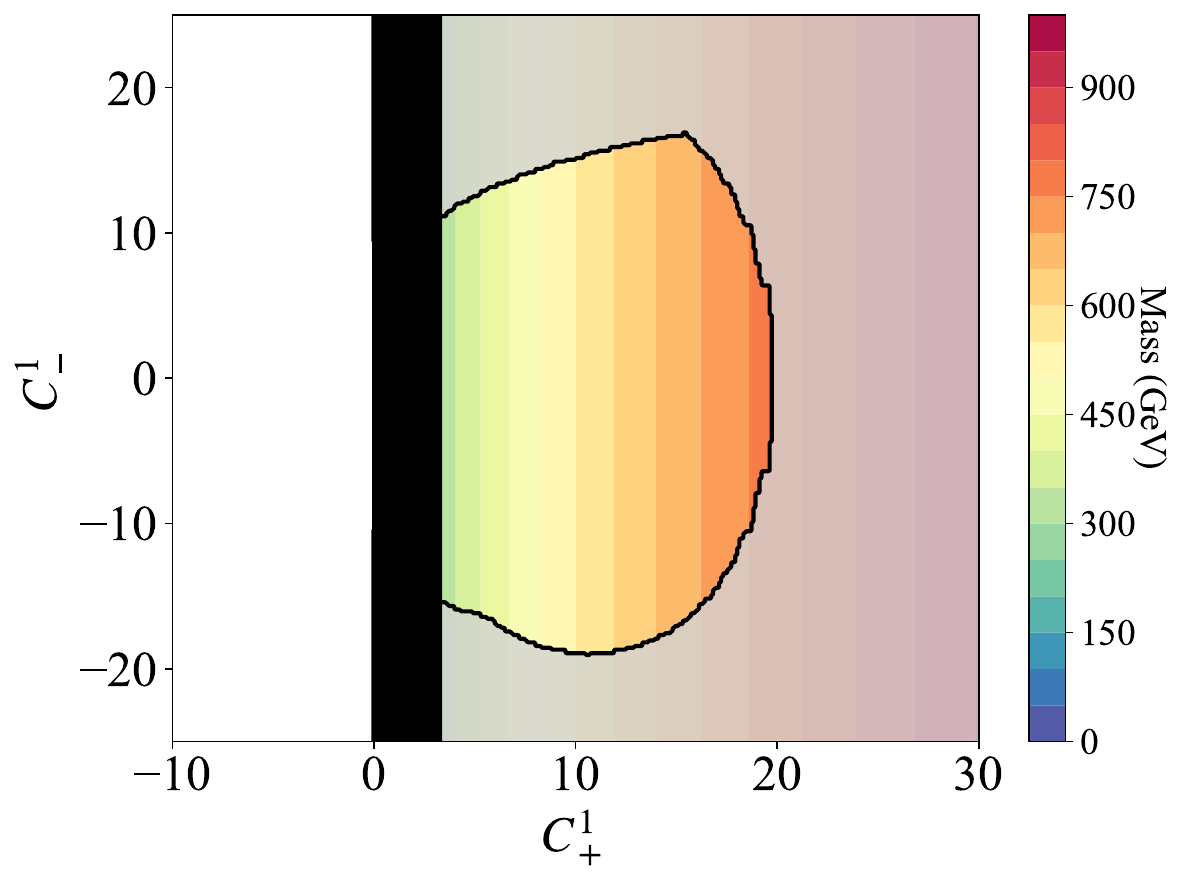} 
  \end{subfigure}
  \caption{
  Constraints for the $[2,2]_0$ Loryon representation in the alignment limit ($\cos(\beta-\alpha)=0$) with $\tan\beta=1$ and fixed mass fractions $f_{0}=f_{1}=0.5$. The left panel varies the singlet couplings $(C_{+}^{0},C_{-}^{0})$ with fixed triplet couplings $(C_{+}^{1},C_{-}^{1})=(5,5)$; as the singlet is neutral, the excluded regions are determined solely by perturbative unitarity (gray) and mass positivity (white, where $m^2 < 0$). The right panel varies the triplet couplings $(C_{+}^{1},C_{-}^{1})$ with fixed singlet couplings $(C_{+}^{0},C_{-}^{0})=(5,5)$; here, the black region indicates additional parameter space excluded by Higgs diphoton decays ($\kappa_\gamma$) due to the charged components of the triplet. The color gradient from blue to red represents the physical Loryon mass within the allowed region.
  }
  \label{fig:tan1cos0}
\end{figure}

\begin{figure}[h!]
  \centering
  \begin{subfigure}[b]{0.49\textwidth}
    \centering
    \includegraphics[width=\textwidth]{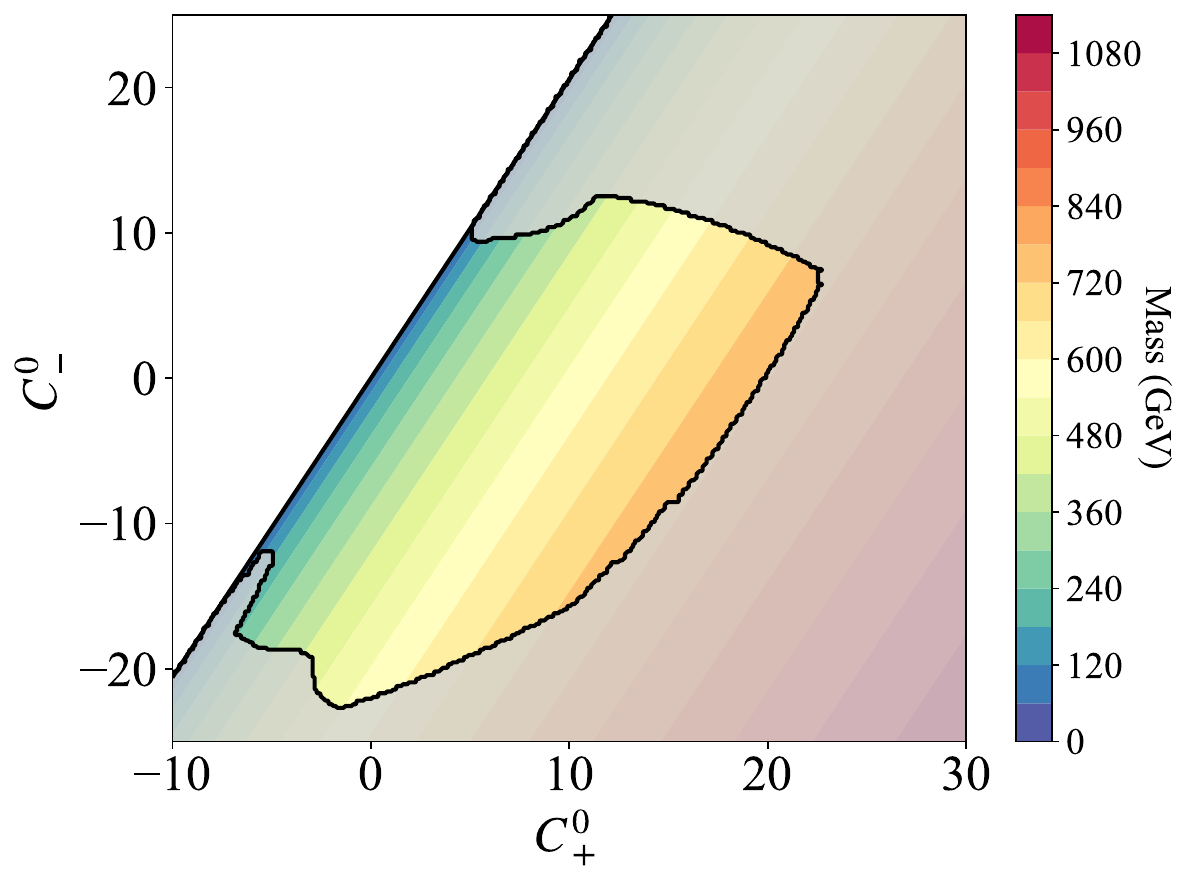} 
  \end{subfigure}
  \hfill
  \begin{subfigure}[b]{0.49\textwidth}
    \centering
    \includegraphics[width=\textwidth]{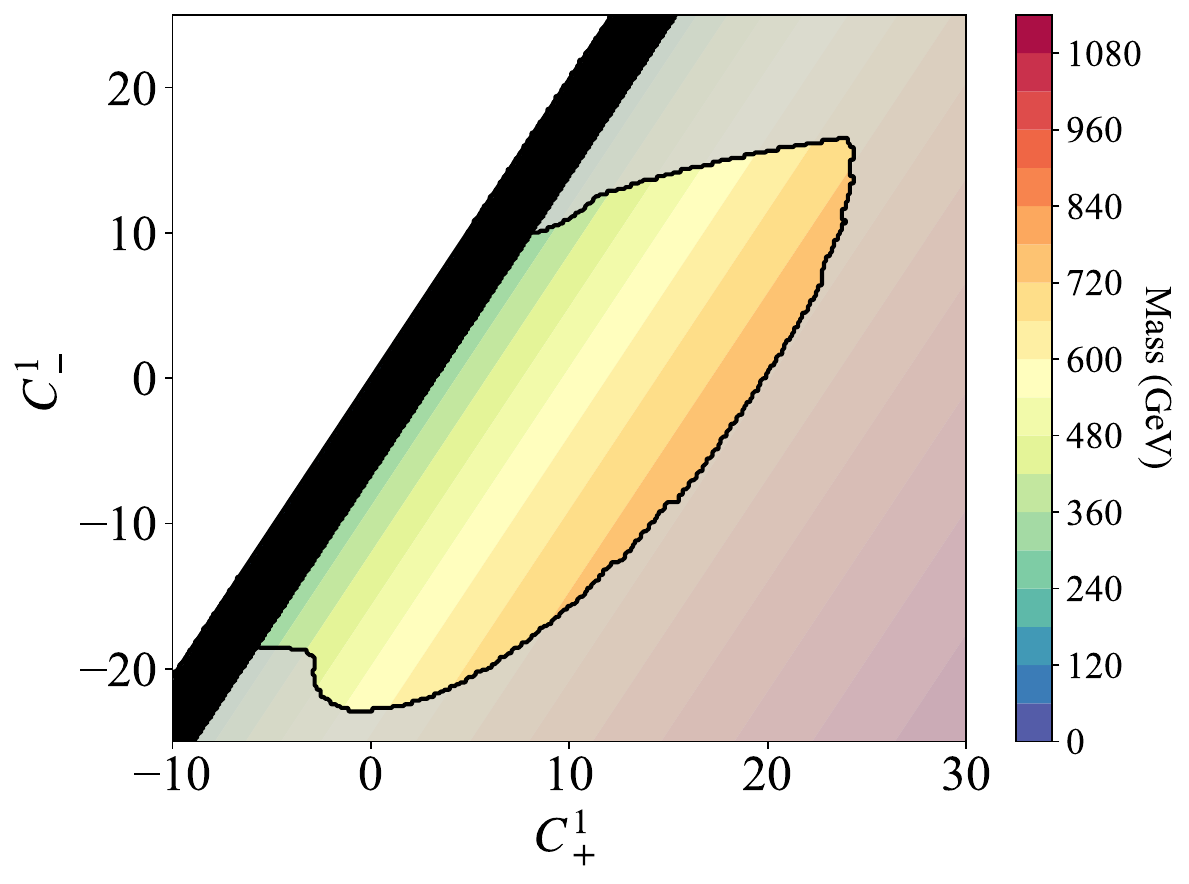} 
  \end{subfigure}
  \caption{
  Similar to figure~\ref{fig:tan1cos0}, with ($\cos(\beta-\alpha)=0$) and $\tan\beta=1.7$. 
  }
  \label{fig:tan1.7cos0}
\end{figure}

\begin{figure}[h!]
  \centering
  \begin{subfigure}[b]{0.49\textwidth}
    \centering
    \includegraphics[width=\textwidth]{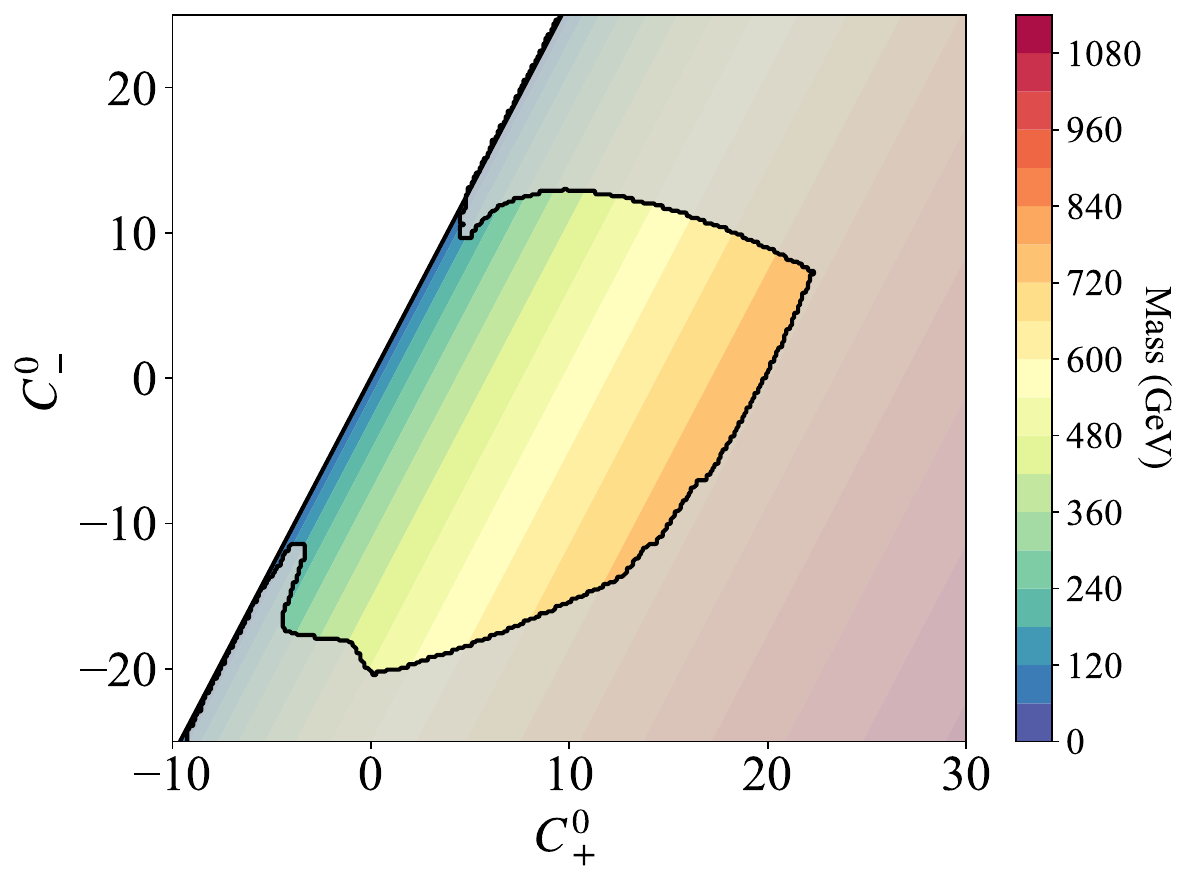} 
  \end{subfigure}
  \hfill
  \begin{subfigure}[b]{0.49\textwidth}
    \centering
    \includegraphics[width=\textwidth]{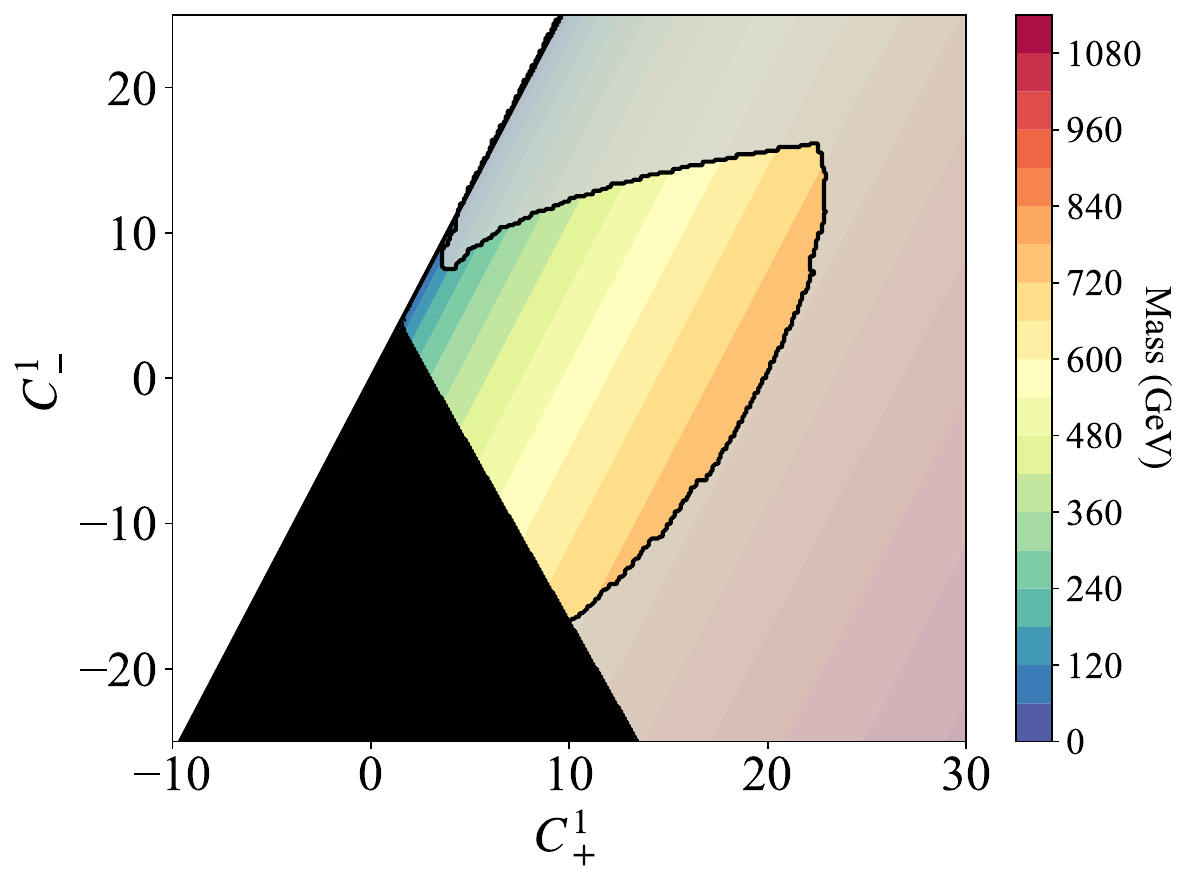} 
  \end{subfigure}
  \caption{
  Similar to figure~\ref{fig:tan1cos0}, with ($\cos(\beta-\alpha)=0.02$) and $\tan\beta=1.5$. 
  }
  \label{fig:tan1.5cos0.02}
\end{figure}

\begin{figure}[h!]
  \centering
  \begin{subfigure}[b]{0.49\textwidth}
    \centering
    \includegraphics[width=\textwidth]{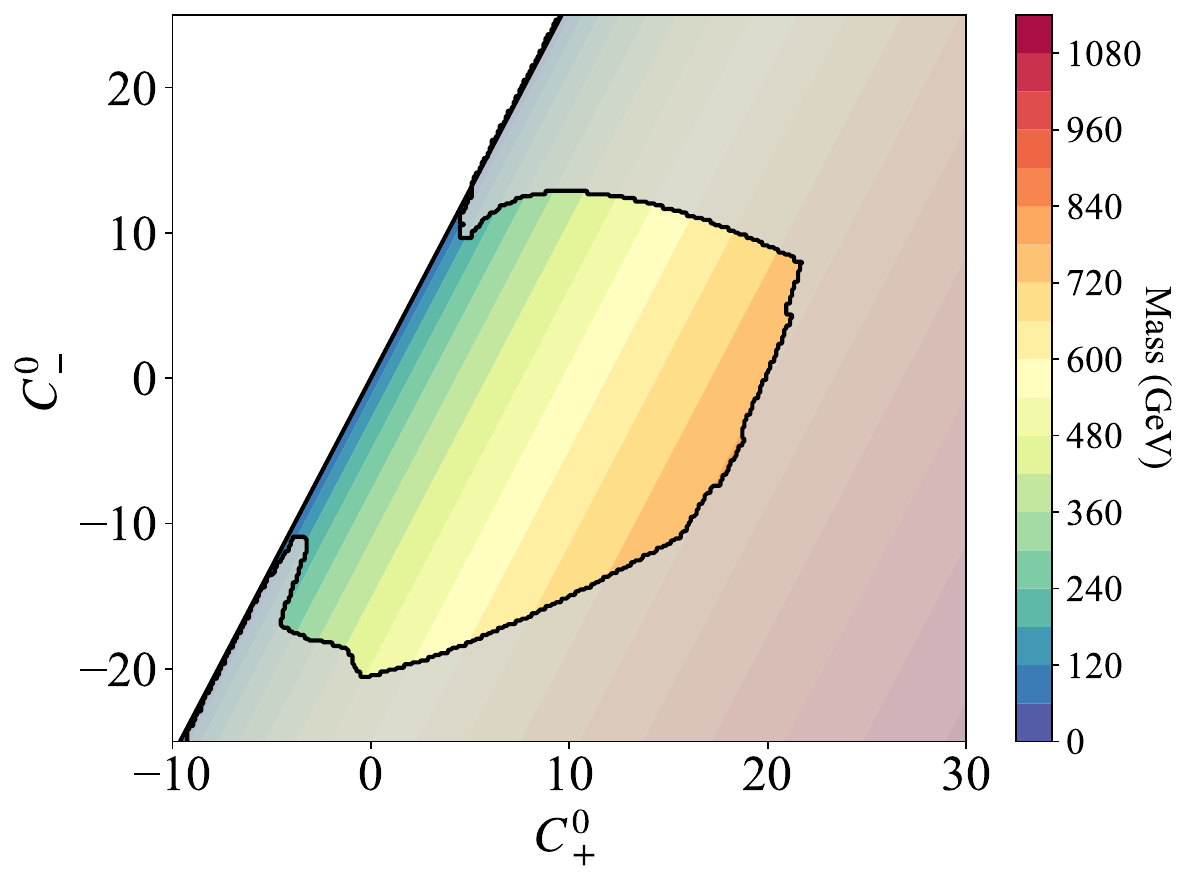} 
  \end{subfigure}
  \hfill
  \begin{subfigure}[b]{0.49\textwidth}
    \centering
    \includegraphics[width=\textwidth]{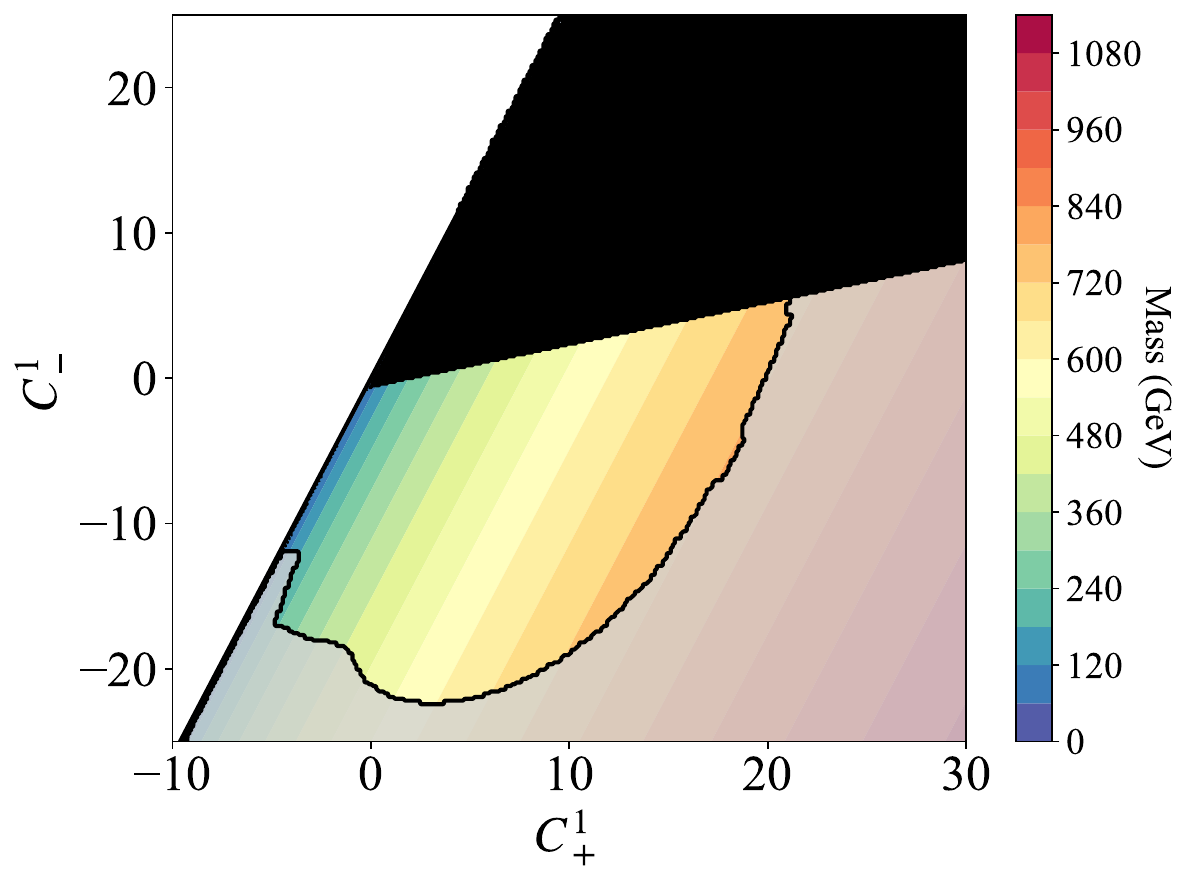}
  \end{subfigure}
  \caption{
  Similar to figure~\ref{fig:tan1cos0}, with ($\cos(\beta-\alpha)=-0.1$) and $\tan\beta=1.5$. 
  }
  \label{fig:tan1.5cos-0.1}
\end{figure}

We present the allowed parameter space for the $[2,2]_0$ Loryon representation in Figures~\ref{fig:tan1cos0} through \ref{fig:tan1.5cos-0.1}. Due to the complexity of the parameter space, which involves four independent couplings ($C_+^0, C_-^0, C_+^1, C_-^1$) defined in Eq.~\eqref{eq:chcH22}, we visualize the constraints by splitting up the dependence on the parameter pairs $C^0_\pm$ and $C^1_\pm$, which affect the properties of the singlet and the triplet, respectively. In the left vs. right panels, we vary one pair of parameters while we fixed the other pair at $(5,5)$ as a benchmark value.

Crucially, we also fix the fractions of mass generated by electroweak symmetry breaking for the singlet and triplet states, denoted as $f_0$ and $f_1$, respectively. The qualitative behavior of the allowed regions follows the logic established in previous sections: maintaining a fixed EWSB mass fraction requires the couplings to grow with the physical mass. Consequently, the allowed parameter space is bounded from above by perturbative unitarity.

A key phenomenological distinction exists between the two multiplets arising from the $[2,2]_0$ representation:
\begin{itemize}
    \item {\bf Singlet scans (varying $C_{\pm}^0$)}: The singlet $s_0$ is electrically neutral and does not contribute to the loop-induced $h \to \gamma\gamma$ decay. Therefore, in the left-hand panels of the figures (where $C_{\pm}^0$ are varied), the exclusion regions are determined solely by perturbative unitarity and mass positivity.
    \item {\bf Triplet scans (varying $C_{\pm}^1$)}: The triplet $s_1$ contains electrically charged scalars. Consequently, the right-hand panels (where $C_{\pm}^1$ are varied) show additional constraints from the Higgs diphoton signal strength $\kappa_\gamma$, indicated by the black shaded regions. 
\end{itemize}

Figures~\ref{fig:tan1cos0} through~\ref{fig:tan1.5cos-0.1} illustrate the constraints for symmetric mass fractions ($f_0 = f_1 = 0.5$) across various 2HDM configurations, ranging from the alignment limit to scenarios with non-zero mixing ($\cos(\beta-\alpha) \neq 0$). The shape of the unitarity boundary is non-trivial; the sharp ``kinks'' observed in the exclusion contours arise from the crossing of eigenvalues in the partial-wave scattering matrix $[a_0]$. As the center-of-mass energy $\sqrt{s}$ varies, different scattering channels dominate the unitarity bound, leading to these distinct inflection points.

\section{Conclusion}
\label{sec:con}

We have considered the scenario of scalar Loryons that acquire a significant fraction of their masses from an extended scalar sector of two Higgs doublets. Both the Loryons and the BSM states in the 2HDM have a non-decoupling nature, which makes the parameter space of this model finite. In order to explore the maximally viable region of parameter space for the Loryons, we adopted a 2HDM benchmark spectrum that is close to the lower end of the experimentally allowed range, as well as benchmark values of $\tan\beta$ and $\cos(\beta - \alpha)$ at and near the alignment limit consistent with experimental constraints. For the smallest few Loryon representations of interest of the custodial symmetry group, we studied bounds from unitarity, Higgs decay measurements, and the absence of Loryon VEVs.

For an uncharged singlet Loryon ($[1,1]_0$) with $f_\phi = 0.5$, masses as high as 700~GeV are still consistent with all the constraints we studied, with the allowed parameter space shrinking as $f_\phi$ increases. For the charged singlet Loryon ($[1,1]_1$), as well as for the triplet Loryon ($[1,3]_0$ or $[3,1]_0$), the parameter space for $f_\phi=0.5$ is more constrained due to Higgs decay, however the allowed mass range for the Loryon remains the same, while almost the entire parameter space is ruled out already for $f_\phi=0.6$. For the singlet-triplet combination ($[2,2]_0$), the unitarity constraints become more complicated due to the high dimensionality of possible scattering channels, however the allowed mass range is in the end similar to the previous cases.

Given the finite region of allowed parameter space, and the strong dependence of the constraints on $f_\phi$, especially when there are charged Loryons present, further improvements in constraining $\kappa_\gamma$, have the potential of ruling out all but the trivial $[1,1]_0$ Loryon. We will explore the impact of direct searches as well as precision Higgs measurements at future colliders on the remaining parameter space in upcoming work.

\acknowledgments{

The authors thank Nathaniel Craig for helpful discussions and feedback.
The research of CK and SM is supported by the National Science Foundation Grant Number PHY-2210562.
TY is funded by the NSF grant PHY-2309456 and the Samsung Science and Technology Foundation under Project Number SSTF-BA2201-06. The research of TY was supported by the Munich Institute for Astro-, Particle and BioPhysics (MIAPbP) which is funded by the Deutsche Forschungsgemeinschaft (DFG, German Research Foundation) under Germany´s Excellence Strategy – EXC-2094 – 390783311.
This work of TY was performed in part at Aspen Center for Physics, which is supported by NSF grant PHY-2210452 and a grant from the Simons Foundation (1161654, Troyer). 
}

\appendix

\section{2HDM Couplings}
\label{app:2hdmfr}

In this appendix, we present the Feynman for the triple and quartic couplings from 2HDM in the physical basis~\cite{Gunion:2002zf}, assuming a $\mathbb{Z}'_2$ symmetry (i.e., $m_{12}^2 = 0$), expressed in terms of the input parameters of the model as in \Eq{eq:2hdmprm}

\subsection*{Cubic Couplings}

\begin{equation}
\makebox[\textwidth][l]{%
  $\begin{aligned}
\begin{tikzpicture}[baseline=(v.base)]
\begin{feynman}
\vertex (v1) {$h$};
\vertex[right=1.3cm of v1] (v); 
\vertex[above right=1cm of v] (v2) {$h$};
\vertex[below right=1cm of v] (v3) {$h$};
\diagram* {[edges=scalar]
(v1) -- (v) -- (v2) ,
(v) -- (v3),
};
\end{feynman}
\end{tikzpicture}
 =\frac{3m_{h}^{2}}{4v}\frac{\cos\b{3\alpha-\beta}+3\cos\b{\alpha+\beta}}{\sin\beta\cos\beta}
  \end{aligned}$%
}
\end{equation}
\begin{equation}
\makebox[\textwidth][l]{%
  $\begin{aligned}
\begin{tikzpicture}[baseline=(v.base)]
\begin{feynman}
\vertex (v1) {$h$};
\vertex[right=1.3cm of v1] (v); 
\vertex[above right=1cm of v] (v2) {$h$};
\vertex[below right=1cm of v] (v3) {$H$};
\diagram* {[edges=scalar]
(v1) -- (v) -- (v2) ,
(v) -- (v3),
};
\end{feynman}
\end{tikzpicture} =\frac{2m_{h}^{2}+m_{H}^{2}}{v}\sin\alpha\cos\alpha\b{\cos\alpha\csc\beta+\sec\beta\sin\alpha}
  \end{aligned}$%
}
\end{equation}
\begin{equation}
\makebox[\textwidth][l]{%
  $\begin{aligned}
\begin{tikzpicture}[baseline=(v.base)]
\begin{feynman}
\vertex (v1) {$h$};
\vertex[right=1.3cm of v1] (v); 
\vertex[above right=1cm of v] (v2) {$H$};
\vertex[below right=1cm of v] (v3) {$H$};
\diagram* {[edges=scalar]
(v1) -- (v) -- (v2) ,
(v) -- (v3),
};
\end{feynman}
\end{tikzpicture} =\frac{m_{h}^{2}+2m_{H}^{2}}{v}\csc2\beta\sin2\alpha\sin\b{\alpha-\beta}
  \end{aligned}$%
}
\end{equation}
\begin{equation}
\makebox[\textwidth][l]{%
  $\begin{aligned}
\begin{tikzpicture}[baseline=(v.base)]
\begin{feynman}
\vertex (v1) {$h$};
\vertex[right=1.3cm of v1] (v); 
\vertex[above right=1cm of v] (v2) {$H^+$};
\vertex[below right=1cm of v] (v3) {$H^-$};
\diagram* {[edges=scalar]
(v1) -- (v) -- [charged scalar](v2) ,
(v) -- [anti charged scalar] (v3),
};
\end{feynman}
\end{tikzpicture} =\frac{1}{v}\s{(m_h^2 + 2m_{H^\pm}^2)\sin(\alpha-\beta) -2m_h^2\cot2\beta\cos(\alpha-\beta) }
  \end{aligned}$%
}
\end{equation}
\begin{equation}
\makebox[\textwidth][l]{%
  $\begin{aligned}
\begin{tikzpicture}[baseline=(v.base)]
\begin{feynman}
\vertex (v1) {$H$};
\vertex[right=1.3cm of v1] (v); 
\vertex[above right=1cm of v] (v2) {$H$};
\vertex[below right=1cm of v] (v3) {$H$};
\diagram* {[edges=scalar]
(v1) -- (v) -- (v2) ,
(v) --  (v3),
};
\end{feynman}
\end{tikzpicture} =\frac{3m_{H}^{2}}{4v}\frac{3\sin\b{\alpha+\beta}-\sin\b{3\alpha-\beta}}{\sin\beta\cos\beta}
  \end{aligned}$%
}
\end{equation}

\subsection*{Quartic Couplings}

\begin{equation}
\makebox[\textwidth][l]{%
  $\begin{aligned}
\begin{tikzpicture}[baseline=(v.base)]
\begin{feynman}
\vertex (v1) {$h$};
\vertex[below right=1.3cm of v1] (v); \vertex[below left=1cm of v] (v4) {$h$}; 
\vertex[above right=1cm of v] (v2) {$h$};
\vertex[below right=1cm of v] (v3) {$h$};
\diagram* {[edges=scalar]
(v1) -- (v) -- (v4) ,
(v2)-- (v) -- (v3),
};
\end{feynman}
\end{tikzpicture} =\frac{3}{16v^{2}}\frac{m_{h}^{2}\s{\cos\b{3\alpha-\beta}+3\cos\b{\alpha+\beta}}^{2}+4m_{H}^{2}\cos^{2}\b{\alpha-\beta}\sin^{2}2\alpha}{\sin^{2}\beta\cos^{2}\beta}
  \end{aligned}$%
}
\end{equation}
\begin{equation}
\makebox[\textwidth][l]{%
  $\begin{aligned}
\begin{tikzpicture}[baseline=(v.base)]
\begin{feynman}
\vertex (v1) {$h$};
\vertex[below right=1.3cm of v1] (v); \vertex[below left=1cm of v] (v4) {$h$}; 
\vertex[above right=1cm of v] (v2) {$h$};
\vertex[below right=1cm of v] (v3) {$H$};
\diagram* {[edges=scalar]
(v1) -- (v) -- (v4) ,
(v2)-- (v) -- (v3),
};
\end{feynman}
\end{tikzpicture} =&\frac{3}{8v^{2}}\cos\b{\alpha-\beta}\csc^{2}\beta\sec^{2}\beta\sin2\alpha \\
&\hskip 3em \times \s{\b{\cos\b{3\alpha-\beta}+3\cos\b{\alpha+\beta}}m_{h}^{2}+2\sin2\alpha\sin\b{\alpha-\beta}m_{H}^{2}}
  \end{aligned}$%
}
\end{equation}
\begin{equation}
\makebox[\textwidth][l]{%
  $\begin{aligned}
\begin{tikzpicture}[baseline=(v.base)]
\begin{feynman}
\vertex (v1) {$h$};
\vertex[below right=1.3cm of v1] (v); \vertex[below left=1cm of v] (v4) {$h$}; 
\vertex[above right=1cm of v] (v2) {$H$};
\vertex[below right=1cm of v] (v3) {$H$};
\diagram* {[edges=scalar]
(v1) -- (v) -- (v4) ,
(v2)-- (v) -- (v3),
};
\end{feynman}
\end{tikzpicture} =&-\frac{1}{4v^{2}}\csc^{2}2\beta\sin2\alpha 
\times[\b{-6\sin2\alpha-3\sin\b{4\alpha-2\beta}+\sin2\beta}m_{h}^{2} \\
&-\b{6\sin2\alpha-3\sin\b{4\alpha-2\beta}+\sin2\beta}m_{H}^{2}]
  \end{aligned}$%
}
\end{equation}
\begin{equation}
\makebox[\textwidth][l]{%
  $\begin{aligned}
\begin{tikzpicture}[baseline=(v.base)]
\begin{feynman}
\vertex (v1) {$h$};
\vertex[below right=1.3cm of v1] (v); \vertex[below left=1cm of v] (v4) {$H$}; 
\vertex[above right=1cm of v] (v2) {$H$};
\vertex[below right=1cm of v] (v3) {$H$};
\diagram* {[edges=scalar]
(v1) -- (v) -- (v4) ,
(v2)-- (v) -- (v3),
};
\end{feynman}
\end{tikzpicture} =&\frac{3}{2v^{2}}\csc^{2}2\beta\sin2\alpha\sin\b{\alpha-\beta} \\
&\hskip2em \times \s{2m_{h}^{2}\cos\b{\alpha-\beta}\sin2\alpha+m_{H}^{2}\b{-\sin\b{3\alpha-\beta}+3\sin\b{\alpha+\beta}}}
  \end{aligned}$%
}
\end{equation}
\begin{equation}
\makebox[\textwidth][l]{%
  $\begin{aligned}
\begin{tikzpicture}[baseline=(v.base)]
\begin{feynman}
\vertex (v1) {$H$};
\vertex[below right=1.3cm of v1] (v); \vertex[below left=1cm of v] (v4) {$H$}; 
\vertex[above right=1cm of v] (v2) {$H$};
\vertex[below right=1cm of v] (v3) {$H$};
\diagram* {[edges=scalar]
(v1) -- (v) -- (v4) ,
(v2)-- (v) -- (v3),
};
\end{feynman}
\end{tikzpicture} =\frac{3}{4v^{2}}\csc^{2}2\beta\s{4m_{h}^{2}\sin^{2}2\alpha\sin^{2}\b{\alpha-\beta}+m_{H}^{2}\b{\sin\b{3\alpha-\beta}-3\sin\b{\alpha+\beta}}^{2}}
  \end{aligned}$%
}
\end{equation}

\section{HEFT Criterion}
\label{app:heft}
In this appendix we provide the details of deriving the effective Lagrangian arising from integrating out the Loryons, when those are heavier than all the 2HDM degrees of freedom. We start with the Lagrangian, 
\begin{equation}
\begin{aligned}
    \mathcal L \supset & \frac{1}{2^\rho}\text{tr}(\partial_\mu \Phi^\dagger \partial^\mu \Phi) -\frac{M^2}{2^\rho} \, \tr(\Phi^\dagger \Phi) \\
&- \frac{A_{ik}}{2^\rho}\, \tr(\Phi^\dagger \Phi)\, \ha\, \tr(\mathcal H^\dagger_i \mathcal H_k) 
- \frac{B_{ik}}{2^\rho}\, 2\,\tr(\Phi^\dagger T_{L}^a \Phi T^{\dot a}_{R})\,2\,\tr(\mathcal H_i^\dagger T_{2}^a \mathcal H_k^{a} T_{2}^{\dot a}).
\label{eq:lagapp}
\end{aligned}
\end{equation}
Next we decompose the $[L,R]$ representation of the custodial group $SU(2)_L\times S(2)_R$ into the representations $V$ 
under the diagonal subgroup $SU(2)_V$. When using matrix notation for a field $\Phi$ under $SU(2)_L\times SU(2)_R$ transformations, we use dotted and undotted indices respectively. When considering transformations under the diagonal subgroup $SU(2)_V$, we denote the components with a subscript of $j,m$ used for angular momentum. The two notations can be related via 
\beq
\Phi_{\mu\dot{\mu}}=\sum_{j,m}\sqrt{\frac{2j+1}{2l+1}}\langle l\mu | jm;r\dot{\mu}\rangle \phi_{jm}
\eeq
where $L=2l+1,R=2r+1,V=2j+1$ and $\langle l\mu | jm;r\dot{\mu}\rangle$ are the Clebsch-Gordon coefficients \cite{Banta:2021dek}. To provide an example, a Higgs doublet expressed as a $2\times 2$ matrix under $SU(2)_L\times SU(2)_R$ transformations can be written as, 
\begin{equation}
    \mathcal{H}=\frac{1}{\sqrt{2}}(h+v)+\frac{i}{\sqrt{2}}\sigma^a\tilde{H}^a
\end{equation}
where $\tilde{H}^a$ is the triplet under the diagonal $SU(2)_V$ symmetry, $h$ is the singlet and $\sigma^a$ are the Pauli matrices. 
 Note that the physical Higgs boson in the SM is the singlet of $SU(2)_V$ while the triplet modes are eaten by the gauge bosons in unitary gauge. In a 2HDM, the two singlets are linear combinations of $h$ and $H$, and the un-eaten triplet contains the CP-odd and charged Higgs modes.

The action formed by the lagrangian above can be expressed in the form
\begin{equation}
    S=i\sum_{j,j',m,m'}\int d^4x\ \phi_{jm}((\partial^2 + m_j^2+\alpha_j)\delta_{jm,j'm'}+\kappa_{jm,j'm'})\phi_{j'm'}
\end{equation}
where we have used integration by parts to bring the derivatives together and defined
\begin{multline}
 \alpha_j=\sum_{i,k,a}A_{\b{ik}}\b{2v_{i}h_{k}+h_{i}h_{k}
 +\tilde{H}_{i}^{a}\tilde{H}_{k}^{a}}\\
 +\b{C_2(l)+C_2(r)-C_2(j)}\sum_{i,k,c}B_{(ik)}\b{2v_{i}h_{k}+h_{i}h_{k}-\tilde{H}_{i}^{c}\tilde{H}_{k}^{c}}
\end{multline}
\begin{multline}
 \kappa_{jm,j'm'}\equiv4\sum_{i,k,a,\dot{a},d,\mu,\nu,\dot{\mu},\dot{\nu}}B_{(ik)} \s{\frac{\sqrt{VV'}}{L}\langle jm;r\dot{\mu}| {l\mu}\rangle\b{T_{L}^{a}}_{\mu\nu}\langle{l\nu}|{j'm';r\dot{\nu}}\rangle\b{T_{R}^{\dot{a}}}_{\dot{\nu}\dot{\mu}}}
 \\
 \times\s{\b{v_{i}+h_{i}}\tilde{H}_{k}^{d}\epsilon^{da\dot{a}}+\tilde{H}_{i}^{a}\tilde{H}_{j}^{\dot{a}}}.
\end{multline}
$m_j$ corresponds to the mass of the Loryon in the $j$ representation,
\begin{equation}
    m_j^2 = M^2 + \frac{1}{2}\b{A_{ij} + B_{ij}[C_2(L)+C_2(R) - C_2(V)]}v_i v_j.
\end{equation}
Note that we have used the symmetrized index convention where $A_{(ij)}\equiv \frac{1}{2}\b{A_{ij}+A_{ji}}$. At this point, we have not imposed the $\mathbb{Z}_2$ symmetry yet and therefore this relation is true even for nonzero $A_{12},A_{21},B_{12}, B_{21}$. 
The one-loop contribution to the effective action  from integrating out the Loryons is then as usual given by \cite{Cohen:2020xca}
\begin{align}
    S&=i\ln \det \b{(\partial^2 + m_j^2+\alpha_j)\delta_{jm,j'm'}+\kappa_{jm,j'm'}}\\
    &=i\text{Tr}\ln \b{(\partial^2 + m_j^2+\alpha_j)\delta_{jm,j'm'}+\kappa_{jm,j'm'}}
\end{align}

Furthermore, as introduced in Ref.\cite{Cohen:2020xca}, we organize the effective action as a derivative expansion, 
 \begin{equation}
     S=S^{(0)}+S^{(2)}+\mathcal{O}(\partial^4)
 \end{equation}
where the leading contribution collects terms with no derivatives acting on $\alpha$ or $\kappa$. Hereafter, we will focus on $S^{(0)}$ only. This is given by

 \begin{align*}
     S^{(0)}\equiv i\sum_{j,m}\int d^4x\int \frac{d^4p}{(2\pi)^4}\ln((-p^2 + m_j^2+\alpha_j)+\kappa_{jm,jm})
 \end{align*}

Subtracting the UV divergence using the $\overline{\text{MS}}$ renormalization scheme, we get
 \begin{multline}
     S^{(0)}=\sum_{j,m,j',m',j'',m''}\int d^4x\frac{1}{16\pi^2}\frac{1}{2}(({m_{j}^{2}+\alpha_{j})}\delta_{jm,j''m''}+\kappa_{jm,j''m''})
     \\
     \times(({m_{j'}^{2}+\alpha_{j'})}\delta_{j''m'',j'm'}+\kappa_{j''m'',j'm'})
     \\
     \ \ \times \b{\ln\mu^{2}\delta_{j'm',jm}-\ln\b{\b{m_{j'}^{2}+\alpha_{j'}}\delta_{j'm',jm}+\kappa_{j'm',jm}}+\frac{3}{2}\delta_{j'm',jm}}
 \end{multline}

\paragraph{The SM limit:}
Let us briefly show how this reproduces the SM result in \cite{Banta:2021dek}. For our purposes, we simply have to go to the limit where the loryon potential is identical to Eq.~\eqref{eq:sl0lag}. In order to reach this limit, we consider taking $A_{i2},B_{i2}\rightarrow 0$. On the other hand, in the unitary gauge, the only triplet that appear in the lagrangian are eaten by the gauge bosons. Therefore, this is equivalent to removing all the Higgses but $h_1$ which we will call $h$ from now, and for notational purposes, replacing $v_1\rightarrow v, v_2\rightarrow 0$ in \Eq{eq:lagapp}. In particular, this gives $\kappa\rightarrow0$ which is the non diagonal element in the action. Similarly, $m_j^2 + \alpha_j\rightarrow M^2 + \ha(A + B\b{C_2(l)+C_2(r)-C_2(j)})(v+h)^2 \equiv M^2 + U_j $ .Therefore, we get
\begin{multline}
S^{(0)}\rightarrow\sum_{j,m,j',m',j'',m''}\int d^4x\frac{1}{16\pi^2}\frac{1}{2}({m_{j}^{2}+\alpha_{j}})\delta_{jm,j''m''}
     \times({m_{j'}^{2}+\alpha_{j'}})\delta_{j''m'',j'm'}
     \\
     \ \ \times \b{\ln\mu^{2}-\ln\b{\b{m_{j'}^{2}+\alpha_{j'}}}+\frac{3}{2}}\delta_{j'm',jm}
 \end{multline}
 The delta functions combine into a single delta function $\delta_{jm,jm}$. But since the expression above is independent of the index $m$, for each $j$, we can take the sum $\sum_m=2j+1=V$. Hence, the above result simplifies to
 \begin{equation}
S^{(0)}\rightarrow\sum_{j}V\int d^4x\frac{1}{16\pi^2}\frac{1}{2}({m_{j}^{2}+\alpha_{j}})^2\times \b{\ln\frac{\mu^{2}}{m_j^2+\alpha_j}+\frac{3}{2}}
 \end{equation}
 which is the result obtained in \Eq{eq:efflagscl} to the zero-derivative order.

Coming back to the 2HDM case, we know that the SMEFT description is only useful for predictions of low energy observables if it converges about the electroweak vacuum. This is controlled by the term inside the logarithm. Expressing the argument inside the logarithm as $M^2+U_j$, the condition then becomes $\frac{\langle U_j\rangle}{M^2}<1\implies \frac{\langle U_j\rangle}{M^2+\langle U_j\rangle}<\frac{1}{2}$ for all the representations $j$, which is in our case
 \begin{equation}
     \frac{\sum_{ik}\ha\b{A_{ik}+\b{C_2(l)+C_2(r)-C_2(j)}B_{ik}}v_iv_k}{m_j^2}<\frac{1}{2}\ \ \  \forall j
 \end{equation}
 
Therefore, we define the HEFT criterion to be 
 \begin{equation}
     \frac{\sum_{ik}\ha[A_{ik}+\b{C_2(l)+C_2(r)-C_2(j)}B_{ik}]v_iv_k}{m_j^2}>\frac{1}{2}\ \ \  \forall j
 \end{equation}
Note that we choose this definition even if the loryon is lighter than the Higgs in which case the above non analyticity condition for the effective lagrangian is invalid.

\clearpage
\bibliographystyle{JHEP}
\bibliography{bibtex}{}
\end{document}